\documentclass[journal]{IEEEtran}

\usepackage{xcolor}
\usepackage{cite}
\usepackage{graphicx}

\usepackage[figuresright]{rotating}
\usepackage[normalem]{ulem}
\useunder{\uline}{\ul}{}
\usepackage{booktabs}
\usepackage{multirow}

\begin{document}

\title{Robust Vocal Quality Feature Embeddings for Dysphonic Voice Detection}

\author{Jianwei~Zhang,~\IEEEmembership{Member,~IEEE,}
        Julie~Liss,~\IEEEmembership{Member,~IEEE,}
        Suren~Jayasuriya,~\IEEEmembership{Member,~IEEE,}
        and~Visar~Berisha,~\IEEEmembership{Member,~IEEE}% <-this % stops a space
\thanks{This manuscript is submitted on July 06, 2022 to IEEE/ACM Transactions on Audio, Speech, and Language Processing for peer-review.}
\thanks{Jianwei Zhang is with School of Electrical, Computer and Energy Engineering, Arizona State University, Tempe, AZ 85281, USA (e-mail: jianwei.zhang@asu.edu).}% <-this % stops a space
}

\maketitle

\begin{abstract}
Approximately 1.2\% of the world's population has impaired voice production. As a result, automatic dysphonic voice detection has attracted considerable academic and clinical interest. However, existing methods for automated voice assessment often fail to generalize outside the training conditions or to other related applications. In this paper, we propose a deep learning framework for generating acoustic feature embeddings sensitive to vocal quality and robust across different corpora. A contrastive loss is combined with a classification loss to train our deep learning model jointly. Data warping methods are used on input voice samples to improve the robustness of our method. Empirical results demonstrate that our method not only achieves high in-corpus and cross-corpus classification accuracy but also generates good embeddings sensitive to voice quality and robust across different corpora. We also compare our results against three baseline methods on clean and three variations of deteriorated in-corpus and cross-corpus datasets and demonstrate that the proposed model consistently outperforms the baseline methods.
\end{abstract}


\IEEEpeerreviewmaketitle

\section{Introduction}

\IEEEPARstart{I}{mpaired} voice production is common across a variety of different underlying conditions. For example, cancerous and non-cancerous masses on the vocal folds can result in aperiodic vibration and atypical harmonic structure~\cite{kuhn2013patient,schultz2011vocal}; several neurological conditions, such as Parkinson’s disease and amyotrophic lateral sclerosis, can lead to hypofunctional dysphonia or strained-strangled voice, respectively~\cite{sewall2006clinical,roth1996spasmodic}; poorly regulated activity of the intrinsic and extrinsic laryngeal muscles results in functional dysphonia~\cite{roy2003functional}. The newest guidelines for diagnosis and assessment of disordered voice recommend the use of laryngoscopy, stroboscopy and endoscopy for the accurate diagnoses of voice disorders~\cite{stachler2018clinical}. The consensus auditory-perceptual evaluation of voice (CAPE-V) is a tool used for clinical auditory-perceptual assessment of dysphonic voice~\cite{zraick2011establishing}. However, these approaches rely on the expertise of clinicians trained to use the specialized equipment or trained on perceptual evaluation and do not scale to evaluating large numbers of individuals. As a result, there have been several recent attempts to assess vocal quality using acoustic features directly extracted from recorded speech.

Several general-purpose acoustic features (such as pitch, harmonic-to-noise ratio, etc.) have been used to characterize dysphonic voice. These features also have been used as input to classical machine learning classifiers for distinguishing between healthy controls and patient groups~\cite{martinez2012voice,souissi2015dimensionality,eskidere2015voice,hemmerling2016voice,verde2018voice,huckvale2021automated}. For example, papers have attempted to diagnose Parkinson disease using general voice features like jitter, shimmer, pitch and wavelet Shannon entropy features~\cite{braga2019automatic,soumaya2021detection}. However, the physiological complexity of the voicing apparatus and its diversity across individuals has challenged the development of acoustic vocal quality features that generalize. Moreover, there is good evidence that these conventional features are easily impacted by confounding factors, such as background noise, recording environments and equipment, recording distance, etc~\cite{ge2021reliable}. Even under the same recording condition, widely-used implementations of these features vary significantly from day to day~\cite{stegmann2020repeatability}. As a result, models trained with these features as input fail to generalize outside of the limited conditions on which they are trained. 

More recently, several studies have proposed using the raw signal, spectrogram, or mel-spectrum as input in deep learning models for assessing vocal quality. Some studies adopt the deep CNN or LSTM structure and train the deep learning models from scratch that classify between dysphonic and healthy voice~\cite{harar2017voice,wu2018deep}. Meanwhile, other studies use pre-trained models from computer vision, such as CaffeNet and ResNet34, and refine the model weights accordingly for dysphonic voice detection~\cite{alhussein2018voice,mohammed2020voice}. However, these models don’t necessarily generalize to other applications and can be difficult to interpret.

In this paper, we propose an acoustic feature embedding that is sensitive to vocal quality. We use contrastive learning in a deep architecture to learn a latent representation that characterizes dysphonic voice, generalizes across different corpora, and is robust to confounding factors. A contrastive loss is used to learn embeddings that represent dysphonia-related latent information. Meanwhile, to ensure the model is not sensitive to latent factors unrelated to the vocal quality, we perform data warping on the training audio samples with different effects, including adding background noise, convolving with environment and device impulse responses, and pitch shifting.

We use the complete Saarbruecken Voice Database (SVD)~\cite{woldert2007saarbruecken} for our training and in-corpus validation and use the Massachusetts Eye and Ear Infirmary Database (MEEI) and Hospital Príncipe de Asturias (HUPA) \cite{arias2011combining} for cross-corpus testing. We do not select a specific dysphonia type subset from all recordings to train the model and instead use all available data to learn the embeddings. We show that our model not only achieves good in-corpus and cross-corpus classification accuracy, but also that the embeddings of voice are consistent across various evaluation datasets. We evaluate the separability of generated embeddings using classification accuracy and adjusted mutual information (AMI) scores~\cite{vinh2010information}. Our empirical results demonstrate that the learned embeddings have good separability and consistency on clean and noisy versions of the test data. We also compare our results against three baseline methods~\cite{harar2017voice,verde2018voice,huckvale2021automated} and demonstrate that the proposed model outperforms them across all experimental conditions.

\section{Related Works}

\subsection{Embeddings for Voice}

An embedding is a relatively low-dimensional representation of a high-dimensional input. Ideally, embeddings capture meaningful characteristics of the input data; and embeddings generated from data that are similar in those meaningful characteristics, are located close together in the embedding space. Good embeddings can be learned once and then reused across different datasets and models. Foundation models in natural language processing (e.g. BERT, GPT-3, etc.) are good examples of this~\cite{devlin2018bert, brown2020language}.

In early studies, researchers first trained a deep learning classification model then accumulated the output activation of the last (or the second to last) hidden linear layer as the embeddings for input features~\cite{variani2014deep,richardson2015deep}. Recently, researchers have proposed finding task-relevant embedding features, which can improve performance on specific tasks, by contrastive representation learning and leveraging labeled data~\cite{tian2020makes}. There are currently several popular contrastive losses in deep learning, such as the triplet loss~\cite{balntas2016learning,li2017deep}, tuple-based end-to-end (TE2E) loss~\cite{heigold2016end}, generalized end-to-end (GE2E) loss~\cite{wan2018generalized}, momentum contrast (MoCo)~\cite{he2020momentum}, and SimCLR~\cite{chen2020simple}. These losses applied to speech and voice signals have resulted in embeddings that capture latent information including speaker identification, speech style, emotion, etc~\cite{variani2014deep,richardson2015deep,balntas2016learning,li2017deep,heigold2016end,wan2018generalized,wang2018style,han2021supervised,lu2019one,charlesworth2021gender,zhou2021seen}.

\subsection{Conventional Features with Classical Machine Learning for Dysphonic Voice Detection}

Different types of speech features have been used in the literature to classify between dysphonic and healthy voice. The CAPE-V uses several perceptual features like roughness and breathiness for evaluating the severity of speech disorders~\cite{zraick2011establishing}. Harar et al. extracted conventional acoustic features (pitch, jitter, shimmer, harmonnic-to-noise ratio, etc.) and mel-frequency cepstral coefficients (MFCC) from sustained phonation recordings for dysphonic voice detection~\cite{harar2018towards}. Other features based on spectro-temporal representations have also been proposed for distinguishing between healthy controls and patients, e.g., interlaced derivative patterns on glottal source excitation~\cite{muhammad2017voice} and wavelet time scattering domain features~\cite{lauraitis2020detection}. Many of these existing features sets have been packaged into open-source speech extraction tools (e.g., OpenSMILE, Praat)~\cite{eyben2010opensmile,praat}. These conventional features have been used as inputs to different classical machine learning models including SVM~\cite{souissi2015dimensionality,huckvale2021automated}, Gaussian mixture models (GMM)~\cite{martinez2012voice,muhammad2017enhanced}, random forests (RF)~\cite{hemmerling2016voice}, Bayesian classification (BC)~\cite{verde2018voice}, and XGBoost~\cite{harar2018towards} to study the acoustic manifestation of different dysphonia types.

However, there is growing evidence that existing feature sets used in clinical speech applications are unreliable. Studies have shown that different recording equipment and environments have large impacts on the extracted acoustic features~\cite{ge2021reliable}. Furthermore, even under consistent recording conditions, many widely-used feature extraction algorithms (e.g., OpenSMILE) do not have good repeatability~\cite{stegmann2020repeatability}. These feature values change from day-to-day even when the clinical condition of the patient does not. These sources of variability (combined with small sample sizes) inevitably lead to overfitting~\cite{berisha2021digital}, such that, even using the same speech features data, the classification accuracy obtained by different classical machine learning methods are very different~\cite{hossain2016healthcare,verde2018voice}.

\subsection{Deep Learning for Dysphonic Voice Detection}

Dysphonic voice detection methods based on deep learning use high-dimensional inputs such as the raw signal~\cite{harar2017voice}, the spectrogram~\cite{wu2018deep}, and MFCCs and their first- and second-order derivatives~\cite{alhussein2018voice,mohammed2020voice}. The idea is that, compared to classical machine learning methods, the additional degrees of freedom available to deep learning models can represent the latent information related to an underlying pathology more accurately. Some of these studies rely on fine-tuning existing pre-trained deep-learning models~\cite{alhussein2018voice,mohammed2020voice}. Other studies have trained deep learning models from scratch. For example, Harar et al. trained convolutional layers in combination with recurrent LSTM layers on raw audio signal~\cite{harar2017voice}. Wu et al. used a deep CNN model for classification and utilized Convolutional Deep Belief Networks (CDBN) for pre-training their deep CNN model to reduce potential overfitting~\cite{wu2018deep}. 

While these papers report impressive performance, it’s unclear how well they generalize outside of the small training as they do not report cross-corpus results. This is likely due to the fact that the models are often trained on a small subset of the large and heterogenous SVD corpus~\cite{muhammad2014pathological,muhammad2017voice,wu2018deep,alhussein2018voice,mohammed2020voice}. Furthermore, the published deep learning methods are not explicitly trained to learn meaningful lower-dimensional embeddings, but instead use a black-box classifier.

In this paper, we combine the virtues of deep learning with the convenience of a conventional set of features that can be used across a variety of applications related to dysphonic voice analysis. A deep model is trained to learn a set of general-purpose vocal quality embeddings. In contrast to most previous studies, the complete SVD database is used for training, and we do not down-select only a subset of homogeneous conditions. Finally, we demonstrate that the embeddings generalize across different corpora.

\section{Methods}

\begin{figure*}[ht]
    \centering
	\includegraphics[width=16cm]{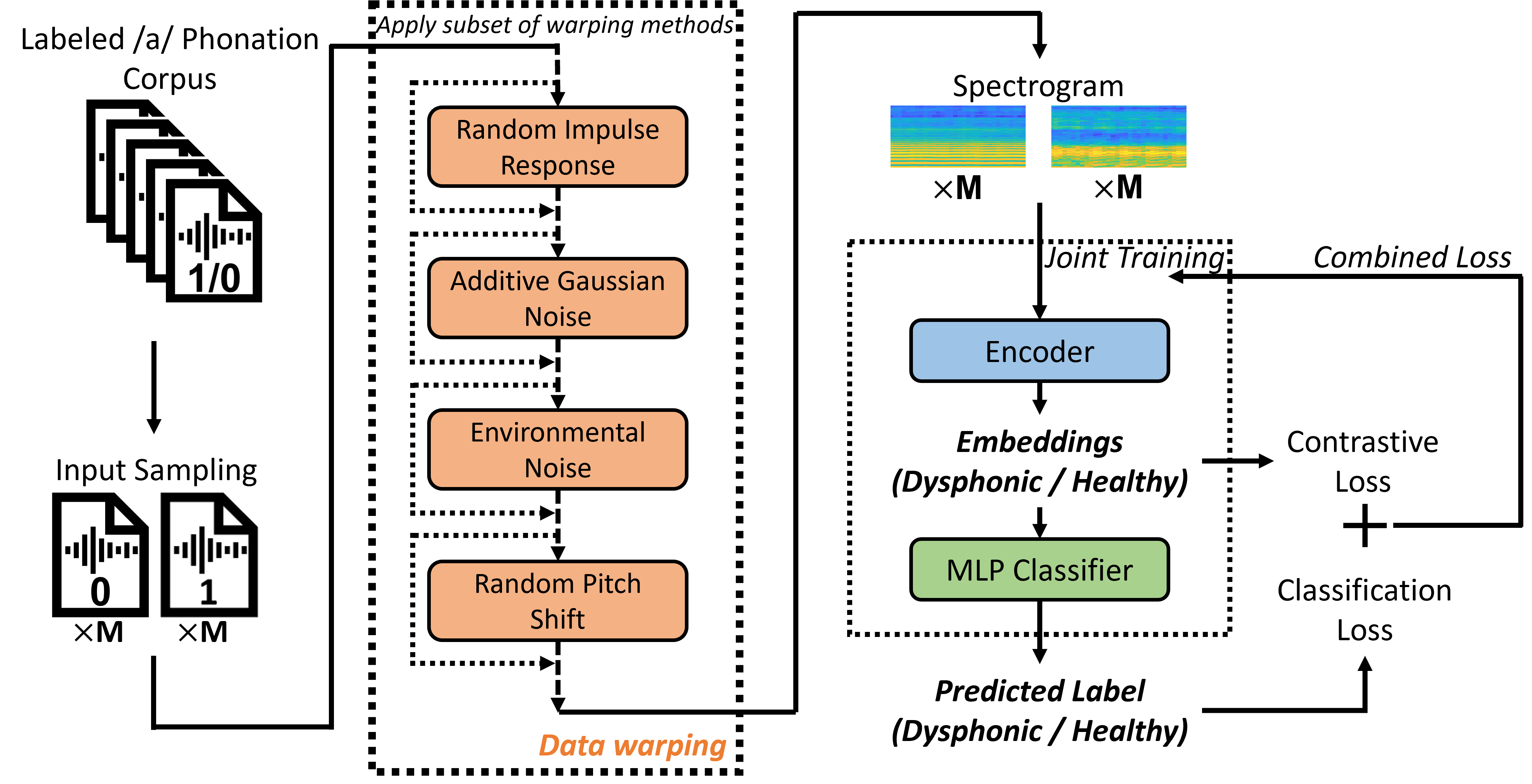}
    \caption{The diagram of our proposed framework. A labeled corpus of sustained /a/ phonation at neutral pitch is used for training. For each batch of training, $M$ healthy and $M$ dysphonic phonation samples are randomly selected from the corpus. These samples are randomly applied with several data warping methods, including convolution with different impulse responses, adding additive Gaussian and environmental noise, and pitch shifting. The warped audio spectrograms are the inputs to our deep learning architecture which contains one encoder and one MLP classifier. We combine a constrastive loss (GE2E) loss, and a classification loss (NLL) to jointly train our deep learning model.}
    \label{fig:design}
\end{figure*}

This paper presents an algorithm for generating speech feature embeddings which are sensitive to voice quality, maximally invariant to nuisance factors, and useful for downstream dysphonic voice classification applications. However, jointly optimizing these goals in practice is difficult. We design a deep learning framework that combines contrastive and classification losses during training: a contrastive loss for generating robust voice embeddings with high separability, and a classification loss for achieving high accuracy in a general-purpose classification problem.

Figure~\ref{fig:design} shows our system diagram including data input, data warping operations, and the network architecture. We use a labelled dysphonic voice corpus for training our model. The sampling procedure for the input randomly selects an equal number of healthy and dysphonic sustained phonation audios of the phoneme /a/ during each training batch. A subset of warping methods including impulse response convolution, additive Gaussian and environmental noise, and pitch shifting are randomly applied to these samples. Applying these data warping methods enables the deep learning model to generalize across different corpora. The warped audio spectrograms are used as input to the deep learning architecture which consists of two modules: an encoder and an MLP classifier. The encoder maps spectrograms to an embedding feature space, and the MLP classifier predicts the label (dysphonic or healthy) based on the embeddings. The contrastive loss is calculated with the labelled embeddings by using a GE2E loss~\cite{wan2018generalized}. We add a classification loss based on the negative log likelihood (NLL) loss to the contrastive loss. Below we describe the details of the network, training, and data warping.

\subsection{Network Architecture}

\begin{figure*}[ht]
    \centering
	\includegraphics[width=14cm]{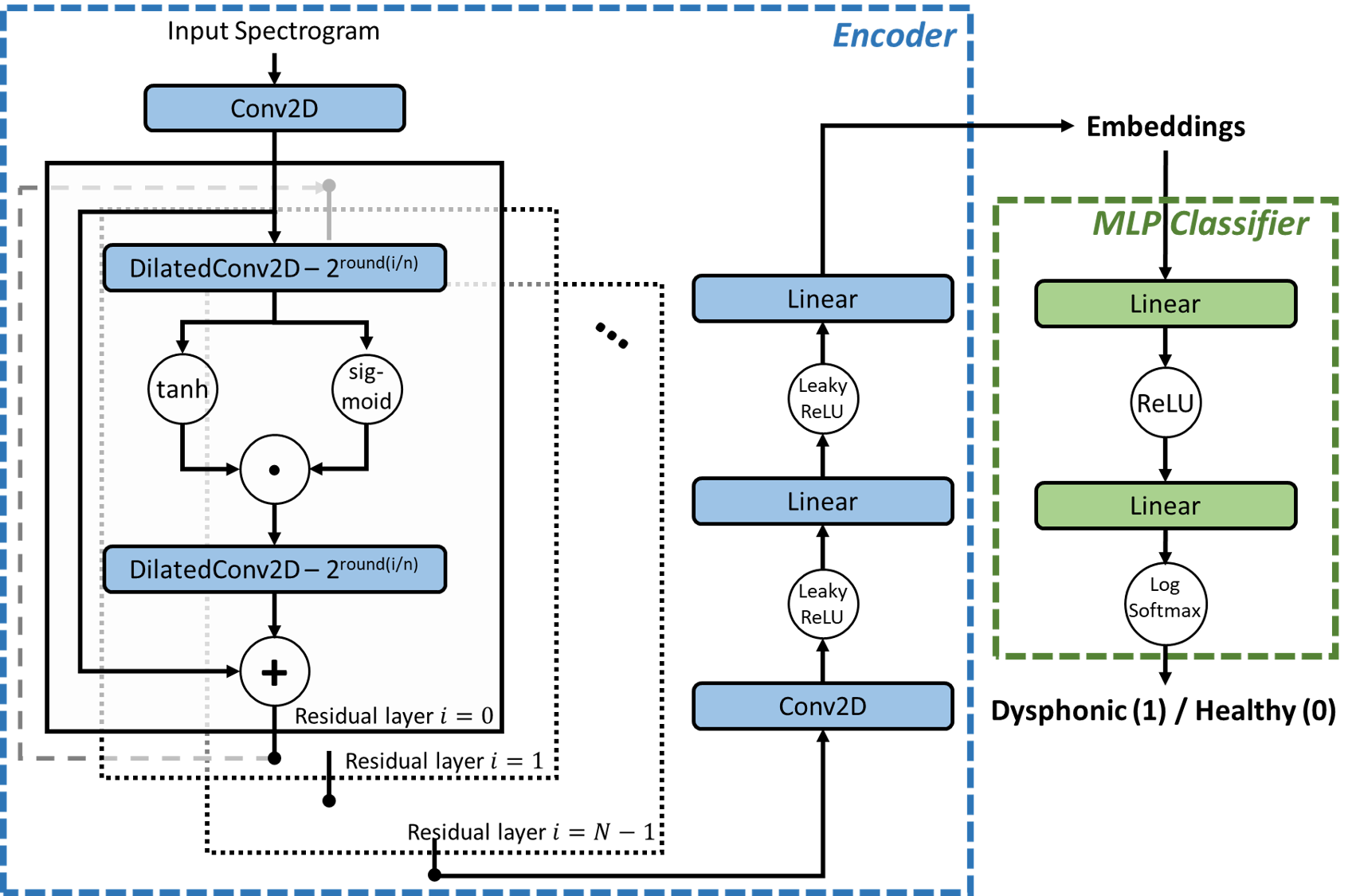}
    \caption{The network architecture of the encoder and MLP classifier. The warped audio spectrograms are passed through several residual blocks in the encoder, and turned into voice embeddings. The MLP classifier predicts the voice label by using the embeddings.}
    \label{fig:model}
\end{figure*}

Figure~\ref{fig:model} shows the network architecture of the encoder and MLP classifier. The encoder is a modified residual network, originally used in DiffWave - a generative speech model~\cite{kong2020diffwave}. The transformed spectrograms pass through an initial 2-D convolution neural network (CNN) layer, and then are processed through residual blocks. The residual network contains $N=15$ layers, and is grouped into $m=5$ blocks with each block containing $n=N/m=3$ layers. We followed the default design of the original paper and conducted additional experiments by varying the blocks number ($m$) selection. We noticed that the model could capture the dysphonic features well when $m \geq 3$. However, the time required for training increases by roughly 50\% when the blocks number ($m$) is increased by 1. We suggest that the blocks number ($m$) should be chosen from 3 to 5. In each residual layer, we use 2 dilated convolution layers with gated-tanh nonlinearities in the middle. The dilation is doubled at each residual group, i.e., $[1,...,2^{n-1}]$. The residual channels number is 32 and the kernel size is 3 for all residual layers. The negative slope is 0.4 for all leaky-ReLU activation layers. The output of the last residual layer is transformed to a 1-D embedding with a shape of $1\times 256$ by using a CNN and linear layers. The MLP classifier simply contains 2 linear layers to project the embeddings from $1\times 256$ to $1\times 2$, and a LogSoftmax activation. The MLP classifier predicts the label of input audio by using the embeddings. For reproducibility, the source code is available online\footnote{Link: https://github.com/vigor-jzhang/dysphonic-emb}.

\subsection{Joint Contrastive and Classification Loss}

Our combined loss function is designed to optimize two objectives: (1) induce separability in embeddings space such that dysphonic samples are near each other and healthy samples are near each other, and (2) achieve high classification accuracy for dysphonic voice detection. For the former, we use a GE2E loss~\cite{wan2018generalized} that was originally used for speaker identification. The GE2E loss enabled their model to produce similar embeddings for different utterances of the same speaker, and dissimilar embeddings for different speakers. We slightly modified the GE2E loss to generate the similar embeddings for different subjects' sustained phonation samples within the same group (either dysphonic or healthy), and then push dysphonic and healthy group centroids as far away as possible.

The GE2E loss uses a scaled cosine similarity between one embedding vector ($\textbf{e}$) to one centroid ($\textbf{c}$):

\begin{equation}
    \mathrm{S}(\textbf{e},\textbf{c}) = \omega\cdot\mathrm{cosine}(\textbf{e},\textbf{c}) + b,
\end{equation}

\noindent where $\omega$ and $b$ are learnable parameters during model training. The contrastive loss for dysphonic embeddings $\textbf{e}_{p,i} (1\leq i\leq M)$ and healthy embeddings $\textbf{e}_{n,i} (1\leq i\leq M)$ are:

\begin{equation}
    L_{C}(\textbf{e}_{p,i}) = 1-\sigma(\mathrm{S}(\textbf{e}_{p,i},\textbf{c}_{p})) + \sigma(\mathrm{S}(\textbf{e}_{p,i},\textbf{c}_{n})),
\end{equation}

\begin{equation}
    L_{C}(\textbf{e}_{n,i}) = 1-\sigma(\mathrm{S}(\textbf{e}_{n,i},\textbf{c}_{n})) + \sigma(\mathrm{S}(\textbf{e}_{n,i},\textbf{c}_{p})),
\end{equation}

\noindent where $\textbf{c}_{p}$ and $\textbf{c}_{n}$ are the centroids of dysphonic and healthy embeddings respectively, and $\sigma(x)=1/(1+\rm{exp}(-x))$ is the sigmoid function. The centroids are calculated as the mean vectors of dysphonic or healthy embeddings for one batch as

\begin{equation}
    \textbf{c}_{p} = \frac{1}{M}\sum^{M}_{i}\textbf{e}_{p,i},
    \textbf{c}_{n} = \frac{1}{M}\sum^{M}_{i}\textbf{e}_{n,i}.
\end{equation}

\noindent The simplified GE2E loss in our case is:

\begin{equation}
    L_{GE2E} = \sum_{i=1}^{M} ( L_{C}(\textbf{e}_{p,i}) + L_{C}(\textbf{e}_{n,i}) ).
\end{equation}

For the classification loss, we use the NLL loss implemented in PyTorch~\cite{paszke2019pytorch}. To jointly optimize the embedder $E(\cdot)$ and the MLP classifier $C(\cdot)$, we combine the GE2E loss and NLL loss,

\begin{equation} \label{eq:loss}
    \mathop{\min}\limits_{E(\cdot),C(\cdot)} L = (1-\lambda) L_{GE2E} + \lambda L_{NLL},
\end{equation}

\noindent where $\lambda$ is a hyperparameter for weighting the relative importance of each loss. In our implementation, the hyperparameter $\lambda$ is set as 0.5 to equally focus on the encoder and MLP classifier training.

\subsection{Data Warping}

The dysphonic voice corpora we used in this paper are collected under clinical settings and quiet environments. If only the raw samples are used for training, the deep learning model performance may degrade when applied to other corpora collected under different conditions. To ensure the model is robust across different corpora and not sensitive to latent factors unrelated to the vocal quality, we perform data warping on the training samples with different effects, including adding background noise, convolution with environment and device impulse responses, and pitch shifting.

We use the DEMAND database for additive environmental background noise sources~\cite{paszke2019pytorch,thiemann2013diverse}. We use three impulse response databases to simulate different environments and devices: the Aachen impulse response database~\cite{jeub2009binaural}, the MIT IR Survey~\cite{traer2016statistics}, and IR data for vintage microphones~\cite{MicIRP}. Half of all impulse response files are randomly selected for training the deep learning model, and others are used for evaluating the robustness and generalizability of our model and baseline methods. We use the Python package, \textit{librosa}, to apply pitch shifting on the input audio samples to avoid the effects of pitch level on the embeddings and classification results~\cite{mcfee2015librosa}.
\section{Implementation}

\subsection{Dataset}

\begin{table}[ht]
\centering
\caption{Speakers number of SVD, MEEI and HUPA databases.}
\label{tab:num}
\begin{tabular}{c|c|c|c|c}
\midrule[1pt]
Dataset & \begin{tabular}[c]{@{}c@{}}Dysphonic\\ Female\end{tabular} & \begin{tabular}[c]{@{}c@{}}Dysphonic\\ Male\end{tabular} & \begin{tabular}[c]{@{}c@{}}Healthy\\ Female\end{tabular} & \begin{tabular}[c]{@{}c@{}}Healthy\\ Male\end{tabular} \\ \midrule[1pt]
SVD     & 727                                                        & 629                                                      & 428                                                      & 259                                                    \\ \hline
MEEI    & 240                                                        & 170                                                      & 32                                                       & 21                                                     \\ \hline
HUPA    & 102                                                        & 67                                                      & 115                                                      & 82                                                    \\ \midrule[1pt]
\end{tabular}
\end{table}

The Saarbruecken Voice Database (SVD) and Massachusetts Eye and Ear Infirmary Database (MEEI) are the two most widely-used open-access disordered speech corpora in dysphonic voice detection studies~\cite{woldert2007saarbruecken}. The MEEI database contains more than 1400 recordings of sustained phonations, which are collected from 53 healthy speakers and 657 speakers diagnosed with different types of dysphonia. The SVD database contains the voice recordings and electroglottography (EGG) from more than 2000 speakers (428 healthy females, 259 healthy males, 727 dysphonic females, 629 dysphonic males). For each voice recording session, there are 13 files: vowels /i,a,u/ with normal, high, low and rising-falling pitch (total of 12 files), and a sentence reading task. We also use the Hospital Príncipe de Asturias (HUPA) corpus\cite{arias2011combining} (shared by the paper authors after request), which contains /a/ sustained phonation recordings of 366 adult Spanish speakers (169 dysphonic and 197 healthy). 

In this paper, we use all speakers in the SVD database as our training and in-corpus validation datasets. All gender-labeled recordings in the MEEI database (410 dysphonic and 53 healthy speakers) are grouped together as one cross-corpus testing dataset. And the complete HUPA dataset is used as a second cross-corpus testing dataset. Table~\ref{tab:num} shows the number of speakers across the different conditions in the SVD, MEEI and HUPA databases. Since the recordings of the sustained phonation /a/ at a habitual pitch exist for all three databases, only these files are used for training, validation, and testing. The baseline methods we compare against also use the same sustained phonations for evaluation. And all the recordings of SVD, MEEI, and HUPA datasets are re-sampled to 25 kHz, the lowest sampling frequency across the three databases.

When splitting the training and in-corpus validation dataset from the SVD database, we ensure the subjects in the training dataset do not appear in the in-corpus validation dataset as some subjects have multiple recording sessions. For each experiment, we perform cross-validation 10 times to characterize the variability in performance. The SVD dataset is split as 70\%/30\% for training and in-corpus validation. As the MEEI and HUPA databases are used as cross-corpus testing datasets, there is no need to split these two corpora.

\subsection{Ablation Studies}

Our proposed method is labeled as \textbf{DW + CLL + CNL} since our method has three major components: data warping (DW), classification loss (CLL) and contrastive loss (CNL). We also explore 5 different ablations of our framework for comparison: 1) \textbf{CLL + CNL} - no data warping, joint use of classification and contrastive losses; 2) \textbf{DW + CLL} - classification loss with data warping; 3) \textbf{DW + CNL} - contrastive loss with data warping; 4) \textbf{CLL} - classification loss only; 5) \textbf{CNL} - contrastive loss only. For the ablations without classification loss, we use an SVM (SVC from \textit{scikit-learn}~\cite{scikit-learn} with default settings) on encoder output embeddings for classification instead of using the MLP classifier.

\subsection{Training and Testing Details}

The input sampling strategy is critical for learning the deep embeddings~\cite{wu2017sampling}. The subjects' genders have a large impact on their voice characteristics and voice embeddings, so we randomly select 16 dysphonic and 16 healthy voices of the same gender for each training batch to reduce this impact.

During training, the selected input audio samples are randomly applied a subset of warping methods before generating the corresponding spectrograms. As shown in the data warping block in Figure~\ref{fig:design}, there is a skip-probability of 0.5 for each warping method (meaning that the probability of no warping applied is 0.0625). When convolving with the impulse responses, one impulse response file is randomly selected for each input sample. For the additive Gaussian and environmental background noise, the energy of noise is randomized from 0 to SNR of 15dB. Finally, when performing pitch shifting, the pitch of the input sample is randomly shifted by -200 to +200 cents.

We use the warped spectrogram with a fixed shape of $24\times 1025$ as the model input. The spectrogram parameters are: window length of 2048, hop size of 512, FFT length of 2048. For each input sample, 24 frames (approximately 0.526 seconds audio) of the spectrogram are randomly selected from the complete spectrogram as the model input.

In experiments, we found that the Pytorch random seed does not have a large effect on the evaluation metrics (random seed experiments results are provided Supplemental Material 1 Section IV) but significantly impacts the shape of visualized results. To compare the visualized results of different ablations, we fixed the random seed as 250 at the beginning when training the model. The SGD optimizer of Pytorch is used for training the model with a learning rate of 0.001 and other default settings. We conducted experiments about learning rate selection and noticed that our model could be trained to converge if the learning rate is smaller than 0.01. For the convenience of storing checkpoints, we use a learning rate of 0.001. We use one NVIDIA Titan Xp graphic card to train our models. We train the proposed and ablation models for 50k steps, which takes approximately 10 hours under our configurations. The MEEI and HUPA testing datasets are unseen during training and validation.

During in-corpus validation and cross-corpus testing, the complete spectrogram of one input sample is used for dysphonic or healthy voice prediction. A sliding 24-frame-window is used to process the entire spectrogram with a 6-frame hop-length, and we average the sliding results: 1) for the ablations with classification loss, the classification predictions of the model are averaged as final results; 2) for the ablations without classification loss, the encoder output embeddings are averaged as final embeddings, and then the SVM is used for final classification prediction. Under our configurations, it takes approximately 5 minutes to generate embeddings and classification predictions of 1,000 3-second phonation recordings.

\subsection{Evaluation Methods}

Since the SVD, MEEI and HUPA datasets are imbalanced, we use balanced accuracy for evaluation:

\begin{equation}
    \mathrm{Balanced~accuracy} = \frac{1}{2}\times(\frac{\mathrm{TP}}{\mathrm{TP}+\mathrm{FN}} + \frac{\mathrm{TN}}{\mathrm{TN}+\mathrm{FP}}).
\end{equation}

\noindent In addition to the classification accuracy, we implement a clustering metric to evaluate the embeddings' separability on dysphonic and healthy voices. The voice embeddings from training, in-corpus validation and cross-corpus testing datasets are first clustered by the $k$-means algorithm~\cite{scikit-learn}. Then the adjusted mutual information (AMI)~\cite{vinh2010information}, is calculated with the ground truth class assignments and the $k$-means clustering algorithm assignments of the same samples. The larger the clustering metric, the better separability of dysphonic and healthy voice embeddings.

We also use a 2-D UMAP projection to visually evaluate the robustness and separability of the embeddings. The UMAP algorithm is a widely-used projection tool for dimensionality reduction~\cite{mcinnes2018umap}. Because the embeddings are in high-dimension space and difficult to visualize, we use UMAP to project the embeddings onto a 2-D plane, where the robustness and separability of the embeddings can be intuitively observed. The same UMAP projection axes are used for the training, in-corpus validation and cross-corpus testing datasets.

\subsection{Baseline Methods}

We rebuilt three baseline methods to compare with our method: (1) P. Harar et al.~\cite{harar2017voice}, which is based on a recurrent convolutional neural network model; (2) L. Verde et al.~\cite{verde2018voice}, which used a conventional features set with different classical machine learning classifiers. We selected the SVM trained by sequential minimum optimization (SMO) as the baseline to compare against; (3) M. Huckvale et al.~\cite{huckvale2021automated}, which uses the ComPare feature set from the OpenSMILE toolkit with the SVM and neural networks methods. We used their model that selects the best 1000 features from 6373 ComPare features to train an SVM classifier as the baseline for comparison. All baseline methods are rebuilt and trained on our data since the original implementations were not available. The three baseline methods are trained and evaluated on our training, in-corpus validation and cross-corpus testing datasets using exactly the same procedures as those outlined in each of the studies. And we run baseline methods with and without the data warping configuration for a fair comparison. For easy reference, we provided the features used in Verde’s and Huckvale’s implementation in Supplementary Material 1 Section II \& III and Supplementary Material 2.

In addition to evaluating the baseline methods by classification accuracy, we also compare the robustness of the embeddings and separability with our method. However, the baseline methods do not directly generate embeddings as our approach does; as a result, we select an intermediate representation learned by each network as a feature embedding. For Harar's method, we treat the 25-dim output of the LSTM layer in their deep learning model as the feature embeddings. For Verde's method, we treat the 1483-dim features for SVM classification as the feature embeddings. For Huckvale's method, we treat the 1000-dim features (the best 1000 features from 6373 ComPare features) for SVM classification as the feature embeddings. Then we use the same evaluation methods, i.e., clustering metric and 2-D UMAP projection, to evaluate the voice embeddings of three baseline methods.
% \usepackage{multirow}
% \usepackage[normalem]{ulem}
\begin{table*}[t]
\centering
\caption{Classification accuracy of our method and baseline methods on the clean and degraded SVD, MEEI and HUPA datasets.}
\label{tab:acc}
\resizebox{2.0\columnwidth}{!}{
\begin{tabular}{c|ccccc|cccc|cccc}
\midrule[1.5pt]
\multirow{2}{*}{}             & \multicolumn{5}{c|}{\textbf{SVD Dataset}}                                                                                                                                                                                                                                                                                                                                                                                                                        & \multicolumn{4}{c|}{\textbf{MEEI Dataset}}                                                                                                                                                                                                                                                                                                                                       & \multicolumn{4}{c}{\textbf{HUPA Dataset}}                                                                                                                                                                                                                                                                                                                                        \\ \cline{2-14} 
                              & \multicolumn{1}{c|}{Train}                                                    & \multicolumn{1}{c|}{IC Clean}                                                                   & \multicolumn{1}{c|}{IC AN}                                                                      & \multicolumn{1}{c|}{IC IR}                                                                      & IC AN+IR                                                                   & \multicolumn{1}{c|}{CC Clean}                                                                   & \multicolumn{1}{c|}{CC AN}                                                                      & \multicolumn{1}{c|}{CC IR}                                                                      & CC AN+IR                                                                   & \multicolumn{1}{c|}{CC Clean}                                                                   & \multicolumn{1}{c|}{CC AN}                                                                      & \multicolumn{1}{c|}{CC IR}                                                                      & CC AN+IR                                                                   \\ \midrule[1.5pt]
{\textcolor{red}{\textbf{DW + CLL + CNL}}} & \multicolumn{1}{c|}{\begin{tabular}[c]{@{}c@{}}80.03\\ ($\pm$1.80)\end{tabular}} & \multicolumn{1}{c|}{\textcolor{red}{\textbf{\begin{tabular}[c]{@{}c@{}}70.77\\ ($\pm$1.05)\end{tabular}}}}          & \multicolumn{1}{c|}{\textcolor{blue}{\textit{\textbf{\begin{tabular}[c]{@{}c@{}}71.23\\ ($\pm$0.95)\end{tabular}}}}} & \multicolumn{1}{c|}{\textcolor{red}{\textbf{\begin{tabular}[c]{@{}c@{}}69.66\\ ($\pm$1.03)\end{tabular}}}}          & \textcolor{red}{\textbf{\begin{tabular}[c]{@{}c@{}}69.21\\ ($\pm$1.50)\end{tabular}}}          & \multicolumn{1}{c|}{\textcolor{blue}{\textit{\textbf{\begin{tabular}[c]{@{}c@{}}82.09\\ ($\pm$1.41)\end{tabular}}}}} & \multicolumn{1}{c|}{\textcolor{red}{\textbf{\begin{tabular}[c]{@{}c@{}}80.92\\ ($\pm$1.17)\end{tabular}}}}          & \multicolumn{1}{c|}{\textcolor{blue}{\textit{\textbf{\begin{tabular}[c]{@{}c@{}}77.42\\ ($\pm$1.37)\end{tabular}}}}} & \textcolor{red}{\textbf{\begin{tabular}[c]{@{}c@{}}81.76\\ ($\pm$0.94)\end{tabular}}}          & \multicolumn{1}{c|}{\textcolor{red}{\textbf{\begin{tabular}[c]{@{}c@{}}66.51\\ ($\pm$0.80)\end{tabular}}}}          & \multicolumn{1}{c|}{\textcolor{blue}{\textit{\textbf{\begin{tabular}[c]{@{}c@{}}65.80\\ ($\pm$1.03)\end{tabular}}}}} & \multicolumn{1}{c|}{\textcolor{red}{\textbf{\begin{tabular}[c]{@{}c@{}}64.76\\ ($\pm$1.08)\end{tabular}}}}          & \textcolor{blue}{\textit{\textbf{\begin{tabular}[c]{@{}c@{}}65.21\\ ($\pm$1.35)\end{tabular}}}} \\ \hline
CLL + CNL                      & \multicolumn{1}{c|}{\begin{tabular}[c]{@{}c@{}}80.59\\ ($\pm$3.00)\end{tabular}} & \multicolumn{1}{c|}{\begin{tabular}[c]{@{}c@{}}67.92\\ ($\pm$0.93)\end{tabular}}                   & \multicolumn{1}{c|}{\textcolor{red}{\textbf{\begin{tabular}[c]{@{}c@{}}72.91\\ ($\pm$2.23)\end{tabular}}}}          & \multicolumn{1}{c|}{\begin{tabular}[c]{@{}c@{}}62.54\\ ($\pm$1.39)\end{tabular}}                   & \begin{tabular}[c]{@{}c@{}}57.68\\ ($\pm$2.30)\end{tabular}                   & \multicolumn{1}{c|}{\begin{tabular}[c]{@{}c@{}}78.22\\ ($\pm$1.74)\end{tabular}}                   & \multicolumn{1}{c|}{\begin{tabular}[c]{@{}c@{}}72.91\\ ($\pm$2.23)\end{tabular}}                   & \multicolumn{1}{c|}{\begin{tabular}[c]{@{}c@{}}70.40\\ ($\pm$1.66)\end{tabular}}                   & \begin{tabular}[c]{@{}c@{}}70.12\\ ($\pm$2.91)\end{tabular}                   & \multicolumn{1}{c|}{\begin{tabular}[c]{@{}c@{}}58.00\\ ($\pm$1.38)\end{tabular}}                   & \multicolumn{1}{c|}{\begin{tabular}[c]{@{}c@{}}52.34\\ ($\pm$0.79)\end{tabular}}                   & \multicolumn{1}{c|}{\begin{tabular}[c]{@{}c@{}}56.74\\ ($\pm$1.86)\end{tabular}}                   & \begin{tabular}[c]{@{}c@{}}52.11\\ ($\pm$0.91)\end{tabular}                   \\ \hline
\textcolor{blue}{\textit{\textbf{DW + CLL}}}    & \multicolumn{1}{c|}{\begin{tabular}[c]{@{}c@{}}76.83\\ ($\pm$0.82)\end{tabular}} & \multicolumn{1}{c|}{\textcolor{blue}{\textit{\textbf{\begin{tabular}[c]{@{}c@{}}70.10\\ ($\pm$0.58)\end{tabular}}}}} & \multicolumn{1}{c|}{\begin{tabular}[c]{@{}c@{}}70.09\\ ($\pm$0.79)\end{tabular}}                   & \multicolumn{1}{c|}{\textcolor{blue}{\textit{\textbf{\begin{tabular}[c]{@{}c@{}}67.30\\ ($\pm$1.35)\end{tabular}}}}} & \textcolor{blue}{\textit{\textbf{\begin{tabular}[c]{@{}c@{}}68.15\\ ($\pm$1.15)\end{tabular}}}} & \multicolumn{1}{c|}{\textcolor{red}{\textbf{\begin{tabular}[c]{@{}c@{}}82.59\\ ($\pm$1.19)\end{tabular}}}}          & \multicolumn{1}{c|}{\textcolor{blue}{\textit{\textbf{\begin{tabular}[c]{@{}c@{}}80.45\\ ($\pm$1.36)\end{tabular}}}}} & \multicolumn{1}{c|}{\textcolor{red}{\textbf{\begin{tabular}[c]{@{}c@{}}78.11\\ ($\pm$2.05)\end{tabular}}}}          & \textcolor{blue}{\textit{\textbf{\begin{tabular}[c]{@{}c@{}}79.75\\ ($\pm$1.85)\end{tabular}}}} & \multicolumn{1}{c|}{\textcolor{blue}{\textit{\textbf{\begin{tabular}[c]{@{}c@{}}66.28\\ ($\pm$0.76)\end{tabular}}}}} & \multicolumn{1}{c|}{\textcolor{red}{\textbf{\begin{tabular}[c]{@{}c@{}}66.18\\ ($\pm$1.13)\end{tabular}}}}          & \multicolumn{1}{c|}{\textcolor{blue}{\textit{\textbf{\begin{tabular}[c]{@{}c@{}}63.56\\ ($\pm$1.52)\end{tabular}}}}} & \textcolor{red}{\textbf{\begin{tabular}[c]{@{}c@{}}65.63\\ ($\pm$1.48)\end{tabular}}} \\ \hline
DW + CNL                     & \multicolumn{1}{c|}{\begin{tabular}[c]{@{}c@{}}87.30\\ ($\pm$1.24)\end{tabular}} & \multicolumn{1}{c|}{\begin{tabular}[c]{@{}c@{}}69.37\\ ($\pm$1.99)\end{tabular}}                   & \multicolumn{1}{c|}{\begin{tabular}[c]{@{}c@{}}64.05\\ ($\pm$1.31)\end{tabular}}                   & \multicolumn{1}{c|}{\begin{tabular}[c]{@{}c@{}}67.25\\ ($\pm$1.30)\end{tabular}}                   & \begin{tabular}[c]{@{}c@{}}64.30\\ ($\pm$1.11)\end{tabular}                   & \multicolumn{1}{c|}{\begin{tabular}[c]{@{}c@{}}74.26\\ ($\pm$1.70)\end{tabular}}                   & \multicolumn{1}{c|}{\begin{tabular}[c]{@{}c@{}}76.39\\ ($\pm$1.61)\end{tabular}}                   & \multicolumn{1}{c|}{\begin{tabular}[c]{@{}c@{}}70.44\\ ($\pm$1.45)\end{tabular}}                   & \begin{tabular}[c]{@{}c@{}}75.71\\ ($\pm$1.93)\end{tabular}                   & \multicolumn{1}{c|}{\begin{tabular}[c]{@{}c@{}}56.45\\ ($\pm$1.58)\end{tabular}}                   & \multicolumn{1}{c|}{\begin{tabular}[c]{@{}c@{}}58.35\\ ($\pm$1.55)\end{tabular}}                   & \multicolumn{1}{c|}{\begin{tabular}[c]{@{}c@{}}57.90\\ ($\pm$1.03)\end{tabular}}                   & \begin{tabular}[c]{@{}c@{}}59.67\\ ($\pm$1.05)\end{tabular}                   \\ \hline
CLL                           & \multicolumn{1}{c|}{\begin{tabular}[c]{@{}c@{}}75.90\\ ($\pm$2.28)\end{tabular}} & \multicolumn{1}{c|}{\begin{tabular}[c]{@{}c@{}}65.10\\ ($\pm$2.92)\end{tabular}}                   & \multicolumn{1}{c|}{\begin{tabular}[c]{@{}c@{}}51.79\\ ($\pm$0.85)\end{tabular}}                   & \multicolumn{1}{c|}{\begin{tabular}[c]{@{}c@{}}59.0\\ ($\pm$1.55)\end{tabular}}                    & \begin{tabular}[c]{@{}c@{}}51.70\\ ($\pm$1.07)\end{tabular}                   & \multicolumn{1}{c|}{\begin{tabular}[c]{@{}c@{}}72.87\\ ($\pm$2.99)\end{tabular}}                   & \multicolumn{1}{c|}{\begin{tabular}[c]{@{}c@{}}57.21\\ ($\pm$3.19)\end{tabular}}                   & \multicolumn{1}{c|}{\begin{tabular}[c]{@{}c@{}}70.05\\ ($\pm$1.15)\end{tabular}}                   & \begin{tabular}[c]{@{}c@{}}57.59\\ ($\pm$4.12)\end{tabular}                   & \multicolumn{1}{c|}{\begin{tabular}[c]{@{}c@{}}57.58\\ ($\pm$1.66)\end{tabular}}                   & \multicolumn{1}{c|}{\begin{tabular}[c]{@{}c@{}}49.97\\ ($\pm$0.43)\end{tabular}}                   & \multicolumn{1}{c|}{\begin{tabular}[c]{@{}c@{}}56.73\\ ($\pm$0.98)\end{tabular}}                   & \begin{tabular}[c]{@{}c@{}}50.09\\ ($\pm$0.17)\end{tabular}                   \\ \hline
CNL                           & \multicolumn{1}{c|}{\begin{tabular}[c]{@{}c@{}}98.07\\ ($\pm$0.25)\end{tabular}} & \multicolumn{1}{c|}{\begin{tabular}[c]{@{}c@{}}67.98\\ ($\pm$1.88)\end{tabular}}                   & \multicolumn{1}{c|}{\begin{tabular}[c]{@{}c@{}}63.64\\ ($\pm$1.02)\end{tabular}}                   & \multicolumn{1}{c|}{\begin{tabular}[c]{@{}c@{}}58.41\\ ($\pm$0.94)\end{tabular}}                   & \begin{tabular}[c]{@{}c@{}}57.57\\ ($\pm$0.99)\end{tabular}                   & \multicolumn{1}{c|}{\begin{tabular}[c]{@{}c@{}}70.83\\ ($\pm$3.36)\end{tabular}}                   & \multicolumn{1}{c|}{\begin{tabular}[c]{@{}c@{}}74.76\\ ($\pm$2.99)\end{tabular}}                   & \multicolumn{1}{c|}{\begin{tabular}[c]{@{}c@{}}61.28\\ ($\pm$1.31)\end{tabular}}                   & \begin{tabular}[c]{@{}c@{}}68.38\\ ($\pm$2.29)\end{tabular}                   & \multicolumn{1}{c|}{\begin{tabular}[c]{@{}c@{}}50.62\\ ($\pm$0.97)\end{tabular}}                   & \multicolumn{1}{c|}{\begin{tabular}[c]{@{}c@{}}49.57\\ ($\pm$0.83)\end{tabular}}                   & \multicolumn{1}{c|}{\begin{tabular}[c]{@{}c@{}}50.72\\ ($\pm$0.81)\end{tabular}}                   & \begin{tabular}[c]{@{}c@{}}51.59\\ ($\pm$1.14)\end{tabular}                   \\ \midrule[1.5pt]
Harar~\cite{harar2017voice} + DW                    & \multicolumn{1}{c|}{\begin{tabular}[c]{@{}c@{}}76.46\\ ($\pm$2.24)\end{tabular}} & \multicolumn{1}{c|}{\begin{tabular}[c]{@{}c@{}}69.14\\ ($\pm$0.92)\end{tabular}}                   & \multicolumn{1}{c|}{\begin{tabular}[c]{@{}c@{}}69.80\\ ($\pm$1.14)\end{tabular}}                   & \multicolumn{1}{c|}{\begin{tabular}[c]{@{}c@{}}64.17\\ ($\pm$1.59)\end{tabular}}                   & \begin{tabular}[c]{@{}c@{}}63.19\\ ($\pm$1.49)\end{tabular}                   & \multicolumn{1}{c|}{\begin{tabular}[c]{@{}c@{}}70.69\\ ($\pm$3.36)\end{tabular}}                   & \multicolumn{1}{c|}{\begin{tabular}[c]{@{}c@{}}71.47\\ ($\pm$3.32)\end{tabular}}                   & \multicolumn{1}{c|}{\begin{tabular}[c]{@{}c@{}}63.16\\ ($\pm$2.23)\end{tabular}}                   & \begin{tabular}[c]{@{}c@{}}67.16\\ ($\pm$1.87)\end{tabular}                   & \multicolumn{1}{c|}{\begin{tabular}[c]{@{}c@{}}56.42\\ ($\pm$2.40)\end{tabular}}                   & \multicolumn{1}{c|}{\begin{tabular}[c]{@{}c@{}}55.33\\ ($\pm$2.34)\end{tabular}}                   & \multicolumn{1}{c|}{\begin{tabular}[c]{@{}c@{}}52.91\\ ($\pm$0.91)\end{tabular}}                   & \begin{tabular}[c]{@{}c@{}}54.00\\ ($\pm$0.65)\end{tabular}                   \\ \hline
Harar~\cite{harar2017voice}                         & \multicolumn{1}{c|}{\begin{tabular}[c]{@{}c@{}}77.42\\ ($\pm$1.66)\end{tabular}} & \multicolumn{1}{c|}{\begin{tabular}[c]{@{}c@{}}68.04\\ ($\pm$1.11)\end{tabular}}                   & \multicolumn{1}{c|}{\begin{tabular}[c]{@{}c@{}}68.93\\ ($\pm$0.90)\end{tabular}}                   & \multicolumn{1}{c|}{\begin{tabular}[c]{@{}c@{}}58.08\\ ($\pm$1.09)\end{tabular}}                   & \begin{tabular}[c]{@{}c@{}}58.10\\ ($\pm$1.51)\end{tabular}                   & \multicolumn{1}{c|}{\begin{tabular}[c]{@{}c@{}}66.14\\ ($\pm$2.43)\end{tabular}}                   & \multicolumn{1}{c|}{\begin{tabular}[c]{@{}c@{}}62.78\\ ($\pm$3.13)\end{tabular}}                   & \multicolumn{1}{c|}{\begin{tabular}[c]{@{}c@{}}63.16\\ ($\pm$2.18)\end{tabular}}                   & \begin{tabular}[c]{@{}c@{}}64.57\\ ($\pm$2.01)\end{tabular}                   & \multicolumn{1}{c|}{\begin{tabular}[c]{@{}c@{}}49.18\\ ($\pm$0.80)\end{tabular}}                   & \multicolumn{1}{c|}{\begin{tabular}[c]{@{}c@{}}49.22\\ ($\pm$0.77)\end{tabular}}                   & \multicolumn{1}{c|}{\begin{tabular}[c]{@{}c@{}}50.04\\ ($\pm$1.54)\end{tabular}}                   & \begin{tabular}[c]{@{}c@{}}50.56\\ ($\pm$0.95)\end{tabular}                   \\ \hline
Verde~\cite{verde2018voice} + DW                    & \multicolumn{1}{c|}{\begin{tabular}[c]{@{}c@{}}87.15\\ ($\pm$1.04)\end{tabular}} & \multicolumn{1}{c|}{\begin{tabular}[c]{@{}c@{}}63.09\\ ($\pm$0.97)\end{tabular}}                   & \multicolumn{1}{c|}{\begin{tabular}[c]{@{}c@{}}62.72\\ ($\pm$1.27)\end{tabular}}                   & \multicolumn{1}{c|}{\begin{tabular}[c]{@{}c@{}}56.67\\ ($\pm$1.06)\end{tabular}}                   & \begin{tabular}[c]{@{}c@{}}59.11\\ ($\pm$1.70)\end{tabular}                   & \multicolumn{1}{c|}{\begin{tabular}[c]{@{}c@{}}71.08\\ ($\pm$1.84)\end{tabular}}                   & \multicolumn{1}{c|}{\begin{tabular}[c]{@{}c@{}}72.15\\ ($\pm$2.02)\end{tabular}}                   & \multicolumn{1}{c|}{\begin{tabular}[c]{@{}c@{}}61.40\\ ($\pm$1.74)\end{tabular}}                   & \begin{tabular}[c]{@{}c@{}}66.33\\ ($\pm$1.56)\end{tabular}                   & \multicolumn{1}{c|}{\begin{tabular}[c]{@{}c@{}}59.83\\ ($\pm$1.44)\end{tabular}}                   & \multicolumn{1}{c|}{\begin{tabular}[c]{@{}c@{}}58.53\\ ($\pm$1.11)\end{tabular}}                   & \multicolumn{1}{c|}{\begin{tabular}[c]{@{}c@{}}54.87\\ ($\pm$1.01)\end{tabular}}                   & \begin{tabular}[c]{@{}c@{}}55.97\\ ($\pm$1.00)\end{tabular}                   \\ \hline
Verde~\cite{verde2018voice}                         & \multicolumn{1}{c|}{\begin{tabular}[c]{@{}c@{}}89.10\\ ($\pm$0.62)\end{tabular}} & \multicolumn{1}{c|}{\begin{tabular}[c]{@{}c@{}}62.74\\ ($\pm$0.91)\end{tabular}}                   & \multicolumn{1}{c|}{\begin{tabular}[c]{@{}c@{}}62.75\\ ($\pm$1.17)\end{tabular}}                   & \multicolumn{1}{c|}{\begin{tabular}[c]{@{}c@{}}53.98\\ ($\pm$0.90)\end{tabular}}                   & \begin{tabular}[c]{@{}c@{}}56.04\\ ($\pm$1.47)\end{tabular}                   & \multicolumn{1}{c|}{\begin{tabular}[c]{@{}c@{}}70.42\\ ($\pm$1.83)\end{tabular}}                   & \multicolumn{1}{c|}{\begin{tabular}[c]{@{}c@{}}70.86\\ ($\pm$1.28)\end{tabular}}                   & \multicolumn{1}{c|}{\begin{tabular}[c]{@{}c@{}}57.06\\ ($\pm$1.07)\end{tabular}}                   & \begin{tabular}[c]{@{}c@{}}61.89\\ ($\pm$0.98)\end{tabular}                   & \multicolumn{1}{c|}{\begin{tabular}[c]{@{}c@{}}59.76\\ ($\pm$1.50)\end{tabular}}                   & \multicolumn{1}{c|}{\begin{tabular}[c]{@{}c@{}}58.46\\ ($\pm$0.97)\end{tabular}}                   & \multicolumn{1}{c|}{\begin{tabular}[c]{@{}c@{}}53.33\\ ($\pm$0.65)\end{tabular}}                   & \begin{tabular}[c]{@{}c@{}}53.39\\ ($\pm$0.53)\end{tabular}                   \\ \hline
Huckvale~\cite{huckvale2021automated} + DW                 & \multicolumn{1}{c|}{\begin{tabular}[c]{@{}c@{}}71.29\\ ($\pm$2.12)\end{tabular}} & \multicolumn{1}{c|}{\begin{tabular}[c]{@{}c@{}}61.04\\ ($\pm$1.45)\end{tabular}}                   & \multicolumn{1}{c|}{\begin{tabular}[c]{@{}c@{}}61.54\\ ($\pm$1.23)\end{tabular}}                   & \multicolumn{1}{c|}{\begin{tabular}[c]{@{}c@{}}55.67\\ ($\pm$1.18)\end{tabular}}                   & \begin{tabular}[c]{@{}c@{}}56.15\\ ($\pm$0.92)\end{tabular}                   & \multicolumn{1}{c|}{\begin{tabular}[c]{@{}c@{}}70.21\\ ($\pm$4.81)\end{tabular}}                   & \multicolumn{1}{c|}{\begin{tabular}[c]{@{}c@{}}71.70\\ ($\pm$4.95)\end{tabular}}                   & \multicolumn{1}{c|}{\begin{tabular}[c]{@{}c@{}}55.07\\ ($\pm$1.80)\end{tabular}}                   & \begin{tabular}[c]{@{}c@{}}59.89\\ ($\pm$2.12)\end{tabular}                   & \multicolumn{1}{c|}{\begin{tabular}[c]{@{}c@{}}55.76\\ ($\pm$1.82)\end{tabular}}                   & \multicolumn{1}{c|}{\begin{tabular}[c]{@{}c@{}}55.81\\ ($\pm$1.73)\end{tabular}}                   & \multicolumn{1}{c|}{\begin{tabular}[c]{@{}c@{}}51.98\\ ($\pm$0.77)\end{tabular}}                   & \begin{tabular}[c]{@{}c@{}}52.57\\ ($\pm$0.80)\end{tabular}                   \\ \hline
Huckvale~\cite{huckvale2021automated}                      & \multicolumn{1}{c|}{\begin{tabular}[c]{@{}c@{}}72.90\\ ($\pm$1.95)\end{tabular}} & \multicolumn{1}{c|}{\begin{tabular}[c]{@{}c@{}}62.55\\ ($\pm$1.22)\end{tabular}}                   & \multicolumn{1}{c|}{\begin{tabular}[c]{@{}c@{}}60.65\\ ($\pm$1.36)\end{tabular}}                   & \multicolumn{1}{c|}{\begin{tabular}[c]{@{}c@{}}54.74\\ ($\pm$1.06)\end{tabular}}                   & \begin{tabular}[c]{@{}c@{}}52.71\\ ($\pm$0.58)\end{tabular}                   & \multicolumn{1}{c|}{\begin{tabular}[c]{@{}c@{}}69.78\\ ($\pm$4.35)\end{tabular}}                   & \multicolumn{1}{c|}{\begin{tabular}[c]{@{}c@{}}68.74\\ ($\pm$4.19)\end{tabular}}                   & \multicolumn{1}{c|}{\begin{tabular}[c]{@{}c@{}}54.60\\ ($\pm$1.08)\end{tabular}}                   & \begin{tabular}[c]{@{}c@{}}55.46\\ ($\pm$1.61)\end{tabular}                   & \multicolumn{1}{c|}{\begin{tabular}[c]{@{}c@{}}54.87\\ ($\pm$1.44)\end{tabular}}                   & \multicolumn{1}{c|}{\begin{tabular}[c]{@{}c@{}}54.16\\ ($\pm$1.32)\end{tabular}}                   & \multicolumn{1}{c|}{\begin{tabular}[c]{@{}c@{}}51.22\\ ($\pm$0.54)\end{tabular}}                   & \begin{tabular}[c]{@{}c@{}}51.41\\ ($\pm$0.47)\end{tabular}                   \\ \midrule[1.5pt]
\end{tabular}
} % resize
\rm{Each result is generated with 10 cross-validations: \textbf{[mean accuracy in \%] (95\% confidence interval in \%)}. \textcolor{red}{\textbf{Red bold}} is the best result across all methods, and \textcolor{blue}{\textit{\textbf{blue bold italic}}} is the second-best result across all methods. \textbf{Abbreviations:} \textbf{DW} - data warping, \textbf{CLL} - classification loss, \textbf{CNL} - contrastive loss; \textbf{IC} - in-corpus, \textbf{CC} - cross-corpus; \textbf{AN} - additive noise, \textbf{IR} - impulse response.}
\end{table*}
% \usepackage{multirow}
% \usepackage[normalem]{ulem}
\begin{table*}[t]
\centering
\caption{Clustering metrics of our method and baseline methods on the clean and degraded SVD, MEEI and HUPA datasets.}
\label{tab:ami}
\resizebox{2.04\columnwidth}{!}{
\begin{tabular}{c|ccccc|cccc|cccc}
\midrule[1.5pt]
\multirow{2}{*}{}                & \multicolumn{5}{c|}{\textbf{SVD Dataset}}                                                                                                                                                                                                                                                                                                                                                                                                                                                    & \multicolumn{4}{c|}{\textbf{MEEI Dataset}}                                                                                                                                                                                                                                                                                                                                               & \multicolumn{4}{c}{\textbf{HUPA Dataset}}                                                                                                                                                                                                                                                                                                                                                \\ \cline{2-14} 
                                 & \multicolumn{1}{c|}{Train}                                                                        & \multicolumn{1}{c|}{IC Clean}                                                                     & \multicolumn{1}{c|}{IC AN}                                                                        & \multicolumn{1}{c|}{IC IR}                                                                        & IC AN+IR                                                                     & \multicolumn{1}{c|}{CC Clean}                                                                     & \multicolumn{1}{c|}{CC AN}                                                                        & \multicolumn{1}{c|}{CC IR}                                                                        & CC AN+IR                                                                     & \multicolumn{1}{c|}{CC Clean}                                                                     & \multicolumn{1}{c|}{CC AN}                                                                        & \multicolumn{1}{c|}{CC IR}                                                                        & CC AN+IR                                                                     \\ \midrule[1.5pt]
\textcolor{blue}{\textit{\textbf{DW + CLL + CNL}}} & \multicolumn{1}{c|}{\textcolor{blue}{\textit{\textbf{\begin{tabular}[c]{@{}c@{}}0.3167\\ ($\pm$0.016)\end{tabular}}}}} & \multicolumn{1}{c|}{\textcolor{blue}{\textit{\textbf{\begin{tabular}[c]{@{}c@{}}0.2647\\ ($\pm$0.017)\end{tabular}}}}} & \multicolumn{1}{c|}{\textcolor{blue}{\textit{\textbf{\begin{tabular}[c]{@{}c@{}}0.2809\\ ($\pm$0.022)\end{tabular}}}}} & \multicolumn{1}{c|}{\textcolor{blue}{\textit{\textbf{\begin{tabular}[c]{@{}c@{}}0.2411\\ ($\pm$0.021)\end{tabular}}}}} & \textcolor{blue}{\textit{\textbf{\begin{tabular}[c]{@{}c@{}}0.2628\\ ($\pm$0.033)\end{tabular}}}} & \multicolumn{1}{c|}{\textcolor{blue}{\textit{\textbf{\begin{tabular}[c]{@{}c@{}}0.1533\\ ($\pm$0.011)\end{tabular}}}}} & \multicolumn{1}{c|}{\textcolor{red}{\textbf{\begin{tabular}[c]{@{}c@{}}0.1551\\ ($\pm$0.014)\end{tabular}}}}          & \multicolumn{1}{c|}{\textcolor{blue}{\textit{\textbf{\begin{tabular}[c]{@{}c@{}}0.1211\\ ($\pm$0.012)\end{tabular}}}}} & \textcolor{red}{\textbf{\begin{tabular}[c]{@{}c@{}}0.1452\\ ($\pm$0.015)\end{tabular}}}          & \multicolumn{1}{c|}{\textcolor{blue}{\textit{\textbf{\begin{tabular}[c]{@{}c@{}}0.1837\\ ($\pm$0.022)\end{tabular}}}}} & \multicolumn{1}{c|}{\textcolor{blue}{\textit{\textbf{\begin{tabular}[c]{@{}c@{}}0.2049\\ ($\pm$0.029)\end{tabular}}}}} & \multicolumn{1}{c|}{\textcolor{blue}{\textit{\textbf{\begin{tabular}[c]{@{}c@{}}0.1621\\ ($\pm$0.025)\end{tabular}}}}} & \textcolor{blue}{\textit{\textbf{\begin{tabular}[c]{@{}c@{}}0.1866\\ ($\pm$0.029)\end{tabular}}}} \\ \hline
CLL + CNL                        & \multicolumn{1}{c|}{\begin{tabular}[c]{@{}c@{}}0.2865\\ ($\pm$0.009)\end{tabular}}                   & \multicolumn{1}{c|}{\begin{tabular}[c]{@{}c@{}}0.1984\\ ($\pm$0.019)\end{tabular}}                   & \multicolumn{1}{c|}{\begin{tabular}[c]{@{}c@{}}0.1980\\ ($\pm$0.028)\end{tabular}}                   & \multicolumn{1}{c|}{\begin{tabular}[c]{@{}c@{}}0.1505\\ ($\pm$0.020)\end{tabular}}                   & \begin{tabular}[c]{@{}c@{}}0.1567\\ ($\pm$0.026)\end{tabular}                   & \multicolumn{1}{c|}{\textcolor{red}{\textbf{\begin{tabular}[c]{@{}c@{}}0.1608\\ ($\pm$0.017)\end{tabular}}}}          & \multicolumn{1}{c|}{\textcolor{blue}{\textit{\textbf{\begin{tabular}[c]{@{}c@{}}0.1550\\ ($\pm$0.018)\end{tabular}}}}} & \multicolumn{1}{c|}{\begin{tabular}[c]{@{}c@{}}0.1072\\ ($\pm$0.016)\end{tabular}}                   & \begin{tabular}[c]{@{}c@{}}0.1309\\ ($\pm$0.019)\end{tabular}                   & \multicolumn{1}{c|}{\begin{tabular}[c]{@{}c@{}}0.1434\\ ($\pm$0.013)\end{tabular}}                   & \multicolumn{1}{c|}{\begin{tabular}[c]{@{}c@{}}0.1396\\ ($\pm$0.017)\end{tabular}}                   & \multicolumn{1}{c|}{\begin{tabular}[c]{@{}c@{}}0.1024\\ ($\pm$0.064)\end{tabular}}                   & \begin{tabular}[c]{@{}c@{}}0.1126\\ ($\pm$0.015)\end{tabular}                   \\ \hline
DW + CLL                         & \multicolumn{1}{c|}{\begin{tabular}[c]{@{}c@{}}0.1243\\ ($\pm$0.009)\end{tabular}}                   & \multicolumn{1}{c|}{\begin{tabular}[c]{@{}c@{}}0.1125\\ ($\pm$0.014)\end{tabular}}                   & \multicolumn{1}{c|}{\begin{tabular}[c]{@{}c@{}}0.0784\\ ($\pm$0.006)\end{tabular}}                   & \multicolumn{1}{c|}{\begin{tabular}[c]{@{}c@{}}0.0500\\ ($\pm$0.009)\end{tabular}}                   & \begin{tabular}[c]{@{}c@{}}0.0653\\ ($\pm$0.005)\end{tabular}                   & \multicolumn{1}{c|}{\begin{tabular}[c]{@{}c@{}}0.0980\\ ($\pm$0.005)\end{tabular}}                   & \multicolumn{1}{c|}{\begin{tabular}[c]{@{}c@{}}0.1077\\ ($\pm$0.003)\end{tabular}}                   & \multicolumn{1}{c|}{\begin{tabular}[c]{@{}c@{}}0.0700\\ ($\pm$0.004)\end{tabular}}                   & \begin{tabular}[c]{@{}c@{}}0.0966\\ ($\pm$0.009)\end{tabular}                   & \multicolumn{1}{c|}{\begin{tabular}[c]{@{}c@{}}0.0967\\ ($\pm$0.009)\end{tabular}}                   & \multicolumn{1}{c|}{\begin{tabular}[c]{@{}c@{}}0.1189\\ ($\pm$0.010)\end{tabular}}                   & \multicolumn{1}{c|}{\begin{tabular}[c]{@{}c@{}}0.0583\\ ($\pm$0.012)\end{tabular}}                   & \begin{tabular}[c]{@{}c@{}}0.0786\\ ($\pm$0.006)\end{tabular}                   \\ \hline
\textcolor{red}{\textbf{DW + CNL}}                & \multicolumn{1}{c|}{\textcolor{red}{\textbf{\begin{tabular}[c]{@{}c@{}}0.3242\\ ($\pm$0.014)\end{tabular}}}}          & \multicolumn{1}{c|}{\textcolor{red}{\textbf{\begin{tabular}[c]{@{}c@{}}0.2856\\ ($\pm$0.012)\end{tabular}}}}          & \multicolumn{1}{c|}{\textcolor{red}{\textbf{\begin{tabular}[c]{@{}c@{}}0.3032\\ ($\pm$0.012)\end{tabular}}}}          & \multicolumn{1}{c|}{\textcolor{red}{\textbf{\begin{tabular}[c]{@{}c@{}}0.2551\\ ($\pm$0.012)\end{tabular}}}}          & \textcolor{red}{\textbf{\begin{tabular}[c]{@{}c@{}}0.2664\\ ($\pm$0.009)\end{tabular}}}          & \multicolumn{1}{c|}{\begin{tabular}[c]{@{}c@{}}0.1421\\ ($\pm$0.013)\end{tabular}}                   & \multicolumn{1}{c|}{\begin{tabular}[c]{@{}c@{}}0.1517\\ ($\pm$0.006)\end{tabular}}                   & \multicolumn{1}{c|}{\textcolor{red}{\textbf{\begin{tabular}[c]{@{}c@{}}0.1268\\ ($\pm$0.009)\end{tabular}}}}          & \textcolor{blue}{\textit{\textbf{\begin{tabular}[c]{@{}c@{}}0.1430\\ ($\pm$0.010)\end{tabular}}}} & \multicolumn{1}{c|}{\textcolor{red}{\textbf{\begin{tabular}[c]{@{}c@{}}0.1968\\ ($\pm$0.014)\end{tabular}}}}          & \multicolumn{1}{c|}{\textcolor{red}{\textbf{\begin{tabular}[c]{@{}c@{}}0.2225\\ ($\pm$0.013)\end{tabular}}}}          & \multicolumn{1}{c|}{\textcolor{red}{\textbf{\begin{tabular}[c]{@{}c@{}}0.1870\\ ($\pm$0.009)\end{tabular}}}}          & \textcolor{red}{\textbf{\begin{tabular}[c]{@{}c@{}}0.2133\\ ($\pm$0.014)\end{tabular}}}          \\ \hline
CLL                              & \multicolumn{1}{c|}{\begin{tabular}[c]{@{}c@{}}0.0740\\ ($\pm$0.007)\end{tabular}}                   & \multicolumn{1}{c|}{\begin{tabular}[c]{@{}c@{}}0.0709\\ ($\pm$0.011)\end{tabular}}                   & \multicolumn{1}{c|}{\begin{tabular}[c]{@{}c@{}}0.0539\\ ($\pm$0.007)\end{tabular}}                   & \multicolumn{1}{c|}{\begin{tabular}[c]{@{}c@{}}0.0300\\ ($\pm$0.007)\end{tabular}}                   & \begin{tabular}[c]{@{}c@{}}0.0307\\ ($\pm$0.005)\end{tabular}                   & \multicolumn{1}{c|}{\begin{tabular}[c]{@{}c@{}}0.0885\\ ($\pm$0.006)\end{tabular}}                   & \multicolumn{1}{c|}{\begin{tabular}[c]{@{}c@{}}0.0973\\ ($\pm$0.005)\end{tabular}}                   & \multicolumn{1}{c|}{\begin{tabular}[c]{@{}c@{}}0.0552\\ ($\pm$0.001)\end{tabular}}                   & \begin{tabular}[c]{@{}c@{}}0.0522\\ ($\pm$0.002)\end{tabular}                   & \multicolumn{1}{c|}{\begin{tabular}[c]{@{}c@{}}0.0498\\ ($\pm$0.005)\end{tabular}}                   & \multicolumn{1}{c|}{\begin{tabular}[c]{@{}c@{}}0.0406\\ ($\pm$0.004)\end{tabular}}                   & \multicolumn{1}{c|}{\begin{tabular}[c]{@{}c@{}}0.0125\\ ($\pm$0.002)\end{tabular}}                   & \begin{tabular}[c]{@{}c@{}}0.0210\\ ($\pm$0.007)\end{tabular}                   \\ \hline
CNL                              & \multicolumn{1}{c|}{\begin{tabular}[c]{@{}c@{}}0.3807\\ ($\pm$0.016)\end{tabular}}                   & \multicolumn{1}{c|}{\begin{tabular}[c]{@{}c@{}}0.0816\\ ($\pm$0.012)\end{tabular}}                   & \multicolumn{1}{c|}{\begin{tabular}[c]{@{}c@{}}0.0999\\ ($\pm$0.016)\end{tabular}}                   & \multicolumn{1}{c|}{\begin{tabular}[c]{@{}c@{}}0.0216\\ ($\pm$0.006)\end{tabular}}                   & \begin{tabular}[c]{@{}c@{}}0.0384\\ ($\pm$0.009)\end{tabular}                   & \multicolumn{1}{c|}{\begin{tabular}[c]{@{}c@{}}0.1020\\ ($\pm$0.013)\end{tabular}}                   & \multicolumn{1}{c|}{\begin{tabular}[c]{@{}c@{}}0.1088\\ ($\pm$0.011)\end{tabular}}                   & \multicolumn{1}{c|}{\begin{tabular}[c]{@{}c@{}}0.0215\\ ($\pm$0.005)\end{tabular}}                   & \begin{tabular}[c]{@{}c@{}}0.0531\\ ($\pm$0.005)\end{tabular}                   & \multicolumn{1}{c|}{\begin{tabular}[c]{@{}c@{}}0.0321\\ ($\pm$0.010)\end{tabular}}                   & \multicolumn{1}{c|}{\begin{tabular}[c]{@{}c@{}}0.0280\\ ($\pm$0.006)\end{tabular}}                   & \multicolumn{1}{c|}{\begin{tabular}[c]{@{}c@{}}0.0163\\ ($\pm$0.006)\end{tabular}}                   & \begin{tabular}[c]{@{}c@{}}0.0205\\ ($\pm$0.006)\end{tabular}                   \\ \midrule[1.5pt]
Harar~\cite{harar2017voice} + DW                       & \multicolumn{1}{c|}{\begin{tabular}[c]{@{}c@{}}0.1491\\ ($\pm$0.018)\end{tabular}}                   & \multicolumn{1}{c|}{\begin{tabular}[c]{@{}c@{}}0.0774\\ ($\pm$0.005)\end{tabular}}                   & \multicolumn{1}{c|}{\begin{tabular}[c]{@{}c@{}}0.0766\\ ($\pm$0.004)\end{tabular}}                   & \multicolumn{1}{c|}{\begin{tabular}[c]{@{}c@{}}0.0495\\ ($\pm$0.008)\end{tabular}}                   & \begin{tabular}[c]{@{}c@{}}0.0476\\ ($\pm$0.012)\end{tabular}                   & \multicolumn{1}{c|}{\begin{tabular}[c]{@{}c@{}}0.0609\\ ($\pm$0.014)\end{tabular}}                   & \multicolumn{1}{c|}{\begin{tabular}[c]{@{}c@{}}0.0610\\ ($\pm$0.014)\end{tabular}}                   & \multicolumn{1}{c|}{\begin{tabular}[c]{@{}c@{}}0.0470\\ ($\pm$0.008)\end{tabular}}                   & \begin{tabular}[c]{@{}c@{}}0.0455\\ ($\pm$0.006)\end{tabular}                   & \multicolumn{1}{c|}{\begin{tabular}[c]{@{}c@{}}0.0300\\ ($\pm$0.015)\end{tabular}}                   & \multicolumn{1}{c|}{\begin{tabular}[c]{@{}c@{}}0.0297\\ ($\pm$0.016)\end{tabular}}                   & \multicolumn{1}{c|}{\begin{tabular}[c]{@{}c@{}}0.0086\\ ($\pm$0.003)\end{tabular}}                   & \begin{tabular}[c]{@{}c@{}}0.0108\\ ($\pm$0.004)\end{tabular}                   \\ \hline
Harar~\cite{harar2017voice}                            & \multicolumn{1}{c|}{\begin{tabular}[c]{@{}c@{}}0.1462\\ ($\pm$0.019)\end{tabular}}                   & \multicolumn{1}{c|}{\begin{tabular}[c]{@{}c@{}}0.0704\\ ($\pm$0.006)\end{tabular}}                   & \multicolumn{1}{c|}{\begin{tabular}[c]{@{}c@{}}0.0697\\ ($\pm$0.005)\end{tabular}}                   & \multicolumn{1}{c|}{\begin{tabular}[c]{@{}c@{}}0.0231\\ ($\pm$0.005)\end{tabular}}                   & \begin{tabular}[c]{@{}c@{}}0.0256\\ ($\pm$0.008)\end{tabular}                   & \multicolumn{1}{c|}{\begin{tabular}[c]{@{}c@{}}0.0422\\ ($\pm$0.014)\end{tabular}}                   & \multicolumn{1}{c|}{\begin{tabular}[c]{@{}c@{}}0.0419\\ ($\pm$0.015)\end{tabular}}                   & \multicolumn{1}{c|}{\begin{tabular}[c]{@{}c@{}}0.0377\\ ($\pm$0.008)\end{tabular}}                   & \begin{tabular}[c]{@{}c@{}}0.0370\\ ($\pm$0.010)\end{tabular}                   & \multicolumn{1}{c|}{\begin{tabular}[c]{@{}c@{}}0.0162\\ ($\pm$0.011)\end{tabular}}                   & \multicolumn{1}{c|}{\begin{tabular}[c]{@{}c@{}}0.0163\\ ($\pm$0.011)\end{tabular}}                   & \multicolumn{1}{c|}{\begin{tabular}[c]{@{}c@{}}0.0087\\ ($\pm$0.007)\end{tabular}}                   & \begin{tabular}[c]{@{}c@{}}0.0089\\ ($\pm$0.005)\end{tabular}                   \\ \hline
Verde~\cite{verde2018voice} + DW                       & \multicolumn{1}{c|}{\begin{tabular}[c]{@{}c@{}}3.4e-4\\ ($\pm$1e-4)\end{tabular}}                    & \multicolumn{1}{c|}{\begin{tabular}[c]{@{}c@{}}0.0015\\ ($\pm$8e-4)\end{tabular}}                    & \multicolumn{1}{c|}{\begin{tabular}[c]{@{}c@{}}0.0022\\ ($\pm$0.001)\end{tabular}}                   & \multicolumn{1}{c|}{\begin{tabular}[c]{@{}c@{}}0.0012\\ ($\pm$0.001)\end{tabular}}                   & \begin{tabular}[c]{@{}c@{}}0.0021\\ ($\pm$0.001)\end{tabular}                   & \multicolumn{1}{c|}{\begin{tabular}[c]{@{}c@{}}0.0230\\ ($\pm$0.005)\end{tabular}}                   & \multicolumn{1}{c|}{\begin{tabular}[c]{@{}c@{}}0.0199\\ ($\pm$0.008)\end{tabular}}                   & \multicolumn{1}{c|}{\begin{tabular}[c]{@{}c@{}}0.0252\\ ($\pm$0.008)\end{tabular}}                   & \begin{tabular}[c]{@{}c@{}}0.0141\\ ($\pm$0.003)\end{tabular}                   & \multicolumn{1}{c|}{\begin{tabular}[c]{@{}c@{}}0.0026\\ ($\pm$8e-4)\end{tabular}}                    & \multicolumn{1}{c|}{\begin{tabular}[c]{@{}c@{}}0.0015\\ ($\pm$1e-4)\end{tabular}}                    & \multicolumn{1}{c|}{\begin{tabular}[c]{@{}c@{}}0.0029\\ ($\pm$3e-4)\end{tabular}}                    & \begin{tabular}[c]{@{}c@{}}0.0031\\ ($\pm$5e-4)\end{tabular}                    \\ \hline
Verde~\cite{verde2018voice}                            & \multicolumn{1}{c|}{\begin{tabular}[c]{@{}c@{}}3.5e-4\\ ($\pm$1e-4)\end{tabular}}                    & \multicolumn{1}{c|}{\begin{tabular}[c]{@{}c@{}}0.0015\\ ($\pm$8e-4)\end{tabular}}                    & \multicolumn{1}{c|}{\begin{tabular}[c]{@{}c@{}}0.0023\\ ($\pm$0.002)\end{tabular}}                   & \multicolumn{1}{c|}{\begin{tabular}[c]{@{}c@{}}0.0013\\ ($\pm$0.001)\end{tabular}}                   & \begin{tabular}[c]{@{}c@{}}0.0025\\ ($\pm$8e-4)\end{tabular}                    & \multicolumn{1}{c|}{\begin{tabular}[c]{@{}c@{}}0.0311\\ ($\pm$0.008)\end{tabular}}                   & \multicolumn{1}{c|}{\begin{tabular}[c]{@{}c@{}}0.0185\\ ($\pm$0.005)\end{tabular}}                   & \multicolumn{1}{c|}{\begin{tabular}[c]{@{}c@{}}0.0255\\ ($\pm$0.007)\end{tabular}}                   & \begin{tabular}[c]{@{}c@{}}0.0153\\ ($\pm$0.004)\end{tabular}                   & \multicolumn{1}{c|}{\begin{tabular}[c]{@{}c@{}}0.0027\\ ($\pm$7e-4)\end{tabular}}                    & \multicolumn{1}{c|}{\begin{tabular}[c]{@{}c@{}}0.0011\\ ($\pm$1e-4)\end{tabular}}                    & \multicolumn{1}{c|}{\begin{tabular}[c]{@{}c@{}}0.0030\\ ($\pm$4e-4)\end{tabular}}                    & \begin{tabular}[c]{@{}c@{}}0.0028\\ ($\pm$5e-4)\end{tabular}                    \\ \hline
Huckvale~\cite{huckvale2021automated} + DW                    & \multicolumn{1}{c|}{\begin{tabular}[c]{@{}c@{}}0.0090\\ ($\pm$0.010)\end{tabular}}                   & \multicolumn{1}{c|}{\begin{tabular}[c]{@{}c@{}}0.0014\\ ($\pm$2e-4)\end{tabular}}                    & \multicolumn{1}{c|}{\begin{tabular}[c]{@{}c@{}}0.0019\\ ($\pm$3e-4)\end{tabular}}                    & \multicolumn{1}{c|}{\begin{tabular}[c]{@{}c@{}}0.0021\\ ($\pm$6e-4)\end{tabular}}                    & \begin{tabular}[c]{@{}c@{}}0.0015\\ ($\pm$5e-4)\end{tabular}                    & \multicolumn{1}{c|}{\begin{tabular}[c]{@{}c@{}}0.0206\\ ($\pm$0.004)\end{tabular}}                   & \multicolumn{1}{c|}{\begin{tabular}[c]{@{}c@{}}0.0165\\ ($\pm$0.006)\end{tabular}}                   & \multicolumn{1}{c|}{\begin{tabular}[c]{@{}c@{}}0.0021\\ ($\pm$6e-4)\end{tabular}}                    & \begin{tabular}[c]{@{}c@{}}0.0133\\ ($\pm$0.006)\end{tabular}                   & \multicolumn{1}{c|}{\begin{tabular}[c]{@{}c@{}}0.0011\\ ($\pm$0.001)\end{tabular}}                   & \multicolumn{1}{c|}{\begin{tabular}[c]{@{}c@{}}0.0063\\ ($\pm$0.004)\end{tabular}}                   & \multicolumn{1}{c|}{\begin{tabular}[c]{@{}c@{}}0.0035\\ ($\pm$0.003)\end{tabular}}                   & \begin{tabular}[c]{@{}c@{}}0.0008\\ ($\pm$6e-4)\end{tabular}                    \\ \hline
Huckvale~\cite{huckvale2021automated}                         & \multicolumn{1}{c|}{\begin{tabular}[c]{@{}c@{}}0.0160\\ ($\pm$0.015)\end{tabular}}                   & \multicolumn{1}{c|}{\begin{tabular}[c]{@{}c@{}}0.0096\\ ($\pm$0.009)\end{tabular}}                   & \multicolumn{1}{c|}{\begin{tabular}[c]{@{}c@{}}0.0014\\ ($\pm$3e-4)\end{tabular}}                    & \multicolumn{1}{c|}{\begin{tabular}[c]{@{}c@{}}0.0020\\ ($\pm$5e-4)\end{tabular}}                    & \begin{tabular}[c]{@{}c@{}}0.0019\\ ($\pm$4e-4)\end{tabular}                    & \multicolumn{1}{c|}{\begin{tabular}[c]{@{}c@{}}0.0215\\ ($\pm$0.003)\end{tabular}}                   & \multicolumn{1}{c|}{\begin{tabular}[c]{@{}c@{}}0.0098\\ ($\pm$0.006)\end{tabular}}                   & \multicolumn{1}{c|}{\begin{tabular}[c]{@{}c@{}}0.0013\\ ($\pm$3e-4)\end{tabular}}                    & \begin{tabular}[c]{@{}c@{}}0.0153\\ ($\pm$0.004)\end{tabular}                   & \multicolumn{1}{c|}{\begin{tabular}[c]{@{}c@{}}5.2e-4\\ ($\pm$2e-5)\end{tabular}}                    & \multicolumn{1}{c|}{\begin{tabular}[c]{@{}c@{}}0.0064\\ ($\pm$0.001)\end{tabular}}                   & \multicolumn{1}{c|}{\begin{tabular}[c]{@{}c@{}}0.0074\\ ($\pm$9e-4)\end{tabular}}                    & \begin{tabular}[c]{@{}c@{}}7.8e-4\\ ($\pm$2e-4)\end{tabular}                    \\ \midrule[1.5pt]
\end{tabular}
} % resize
\rm{Each result is generated with 10 cross-validations: \textbf{[mean clustering metric] (95\% confidence interval)}. \textcolor{red}{\textbf{Red bold}} is the best result across all methods, and \textcolor{blue}{\textit{\textbf{blue bold italic}}} is the second-best result across all methods. \textbf{Abbreviations:} \textbf{DW} - data warping, \textbf{CLL} - classification loss, \textbf{CNL} - contrastive loss; \textbf{IC} - in-corpus, \textbf{CC} - cross-corpus; \textbf{AN} - additive noise, \textbf{IR} - impulse response.}
\end{table*}

\section{Experimental Results}

\begin{figure*}[htbp]
    \centering
	\includegraphics[width=14cm]{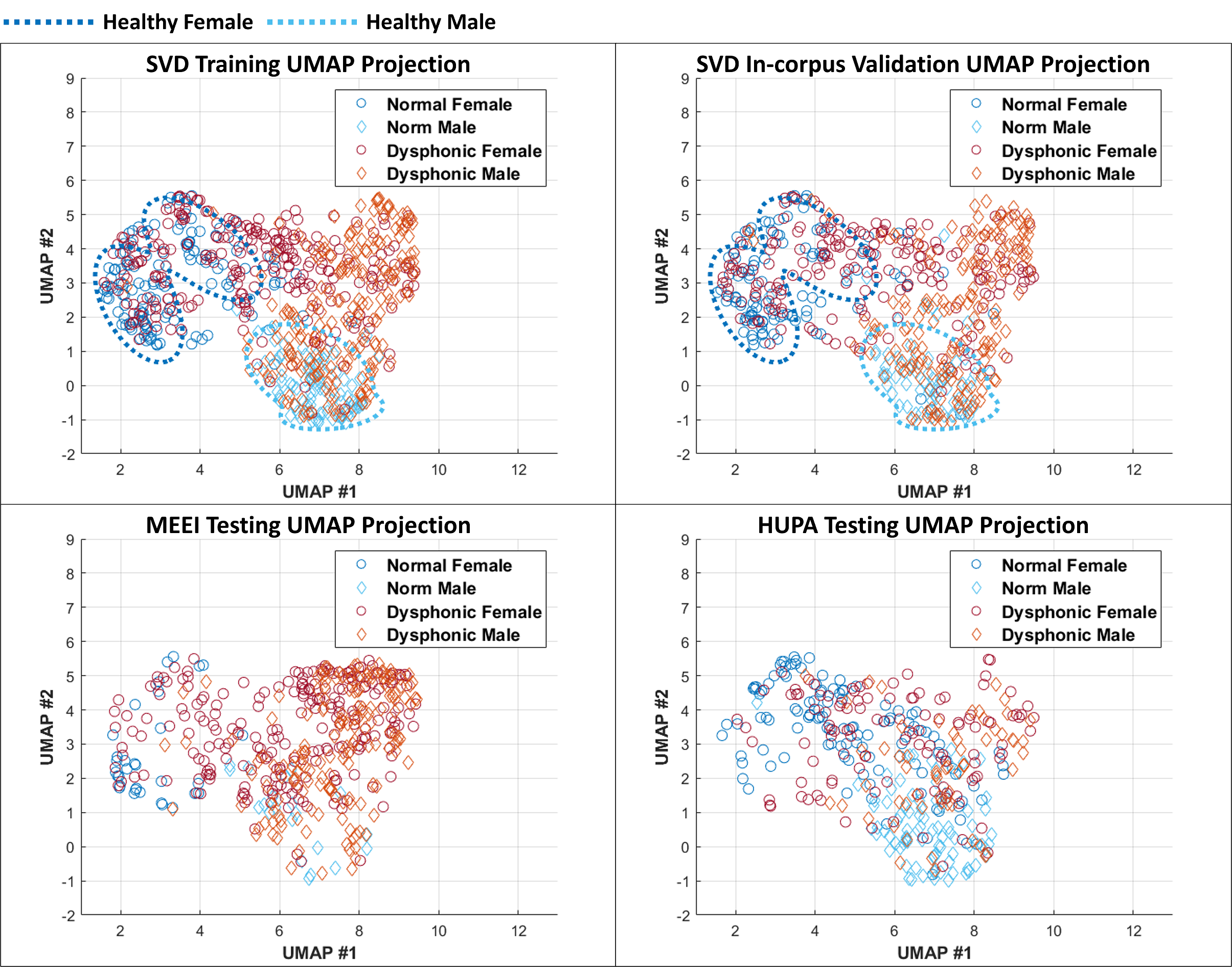}
    \caption{UMAP projection of embeddings generated by our method. The borderline is drawn by applying a Gaussian mixture model on UMAP reduced data, with logscale probability score set to $-3.5$.}
    \label{fig:our}
\end{figure*}

\subsection{Performance of Our Method}

Our proposed deep learning framework achieves a good balance on our two targets: (1) induce separability in the embeddings feature space between the two groups, and (2) high classification accuracy for dysphonic voice detection. Tables \ref{tab:acc} and \ref{tab:ami} provides the classification accuracy and cluster metric results for our methods. We also provide the confusion matrix of 1 of the 10 cross-validation experiments for each method on the clean SVD training, in-corpus validation, MEEI and HUPA cross-corpus testing datasets in Supplementary Material 1 Section I. Each result in the tables is generated by running 10 cross-validations, and a 95\% confidence interval is provided along with the mean value of classification accuracy or clustering metric. Our proposed method achieves good classification accuracy on in-corpus validation 70.77 ($\pm$1.05)\%, MEEI cross-corpus testing 82.09 ($\pm$1.41)\%, and HUPA cross-corpus testing 66.51 ($\pm$0.80)\%. Furthermore, the proposed method outperforms baseline methods on the clustering metrics, which are 0.3167 ($\pm$0.016), 0.2647 ($\pm$0.017), 0.1533 ($\pm$0.011), and 0.1837 ($\pm$0.022) for training, in-corpus validation, MEEI cross-corpus testing and HUPA cross-corpus testing respectively.

We also run an experiment of model fine-tuning on the HUPA dataset, which is a common practice in real deep learning model implementation. We perform cross-validation 10 times for this experiment to characterize performance. The dataset splitting ratio is 70\%/30\%. With the same configurations described in Section IV.A, the model achieves a classification accuracy of 74.49 ($\pm$1.42)\% HUPA testing dataset after 2.5k fine-tuning steps. As a comparison, the original HUPA publication achieved classification accuracy of 76.61 ($\pm$4.30)\% with perturbation features and 69.62 ($\pm$4.67)\% with MFCC features.

As stated in Section IV.B, we also explore 5 different ablations of our framework for comparison. Compared with the proposed method, these configurations do not generate good embeddings and achieve high classification accuracy at the same time. In Table \ref{tab:acc}, although \textbf{DW + CLL} is the second-best configuration among the proposed method and 5 ablations, its clustering metric results are low as shown in Table \ref{tab:ami}. Similarly, \textbf{DW + CNL} achieves the best clustering metrics among the proposed method and 5 ablations, but its classification accuracy results on cross-corpus testing are low. Our experiments demonstrate that data warping is also essential for model generalizability. Comparing the performance of \textbf{DW + CLL + CNL} and \textbf{CLL + CNL} on the HUPA dataset, the classification accuracy is 66.51 ($\pm$0.80)\% vs. 58.00 ($\pm$1.38)\%, and clustering metric is 0.1837 ($\pm$0.022) vs. 0.1434 ($\pm$0.013) respectively. The results demonstrate that the absence of data warping leads to a significant decrease in generalizability. And the accuracy and clustering metric comparison results of other configurations with/without data warping also provide evidence of the same. The contrastive loss, classification loss, and data warping are important to achieve good performance across the different metrics based on the empirical results.

\begin{figure*}[htbp]
    \centering
	\includegraphics[width=14.5cm]{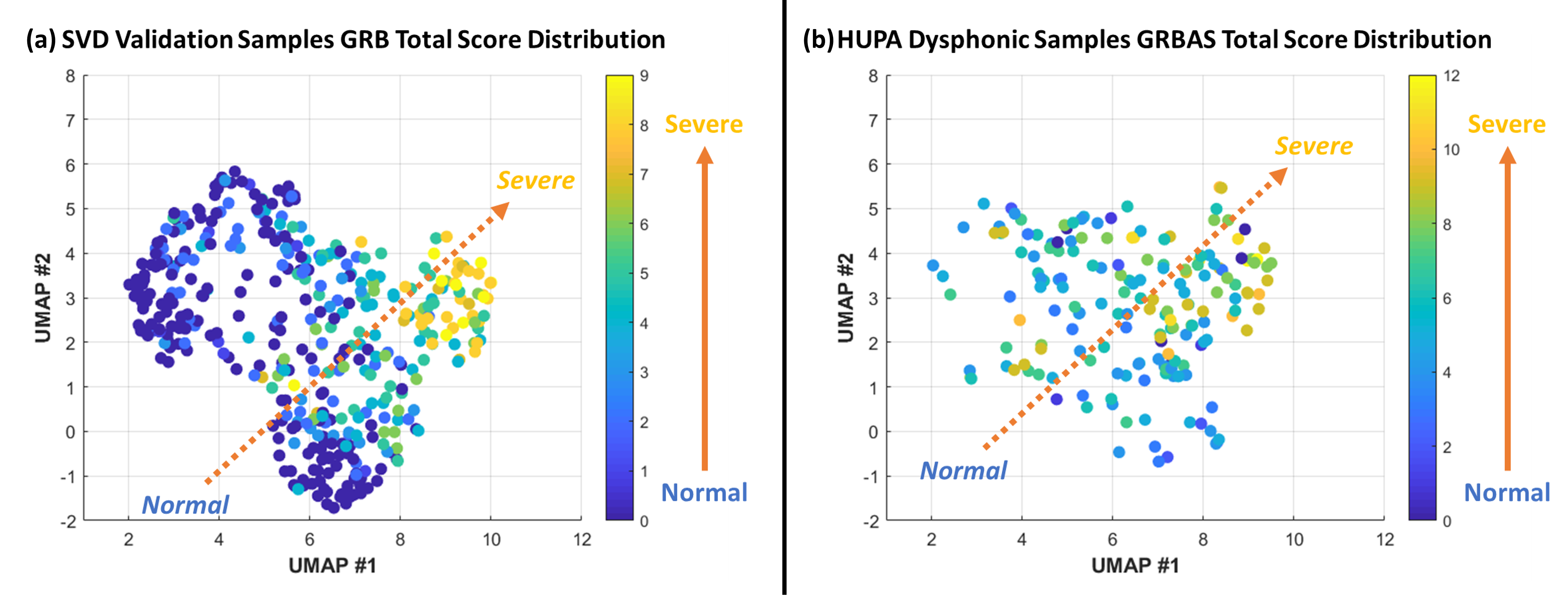}
    \caption{(a) The embeddings UMAP projection of SVD validation samples with GRB total score. (b) The embeddings UMAP projection of dysphonic samples in HUPA testing dataset with GRBAS total score. The point color shows the GRB(AS) total scores, where the color from blue to yellow means the dysphonic severity is from normal to severe. The severity of the dysphonic samples are increasing from the bottom-left to the top-right.}
    \label{fig:hupa_grbas}
\end{figure*}

In addition to the classification accuracy and clustering metrics results, the 2-D UMAP projections also show that the proposed method can produce voice embeddings that cluster based on vocal quality and are robust across different corpora. Figure~\ref{fig:our} shows generated embeddings of training, in-corpus validation, MEEI, and HUPA datasets. The healthy female and male samples are concentrated on the top-left and bottom-right respectively. The dysphonic female and male samples are located on top and right respectively. Based on this figure, a reasonable hypothesis is that the severity of the embedded speech samples increases from the bottom-left to the top-right. This is supported by the UMAP projection of the SVD validation samples with GRB scores and the HUPA dataset with GRBAS scores. GRBAS is a perceptual evaluation scale for rating severity of dysphonia~\cite{hirano1981clinical}. The categorical traits of the GRBAS scale are Grade (G), Roughness (R), Breathiness (B), Asthenia (A), and Strain (S). There is also a simplified scale called GRB that is limited to the G, R, and B dimensions. Arias-Londono et al. provide GRB evaluation for a subset of the SVD dataset~\cite{arias2019multimodal}, and the HUPA dataset contains the GRBAS scores for dysphonic samples. As shown in Figure~\ref{fig:hupa_grbas}, the severity of the dysphonic samples is increasing from the bottom-left to the top-right, and most severe voice samples are located at top-right. These UMAP embeddings provide evidence that our method can capture the latent information related to vocal quality, and the embeddings are consistent across different corpora.

An interactive version of the UMAP projection of the in-corpus validation dataset is provided online\footnote{Link: https://vigor-jzhang.github.io/dysphonic-emb-interactive}. This material contains the raw audio of some subjects and can be played by clicking the individual points on the figure. We strongly recommend that readers listen to the raw audio samples provided. In general, voices located in the top-right corner present with marked dysphonia. Voice samples located on the top-left and bottom-right present with healthy voice quality or mild dysphonia. We draw the reader's attention to two axes on the embeddings distribution: embeddings (1) from the bottom-left to the top-right and (2) from the bottom-left to the top-left which are related to (1) vocal quality and (2) voice pitch respectively. These two axes are also consistent with our training design: (1) a contrastive loss is used to separate dysphonic and healthy voice, and (2) samples either from male or female subjects are selected in one training batch. This result further illustrates the proposed acoustic feature embeddings are sensitive to the vocal quality and voice characteristics.

The performance results and UMAP projections across the three corpora provide strong evidence for the generalizability of our proposed approach. The slightly different in the performance and UMAP projection results could be due to several reasons: the dysphonia subtypes across the various databases are different; SVD, MEEI, and HUPA contain native speakers of different languages (German English and Spanish, respectively), and there may be subtle influences of native language on phonation acoustics; and these datasets do not have the same dysphonic voice types and severity distributions, data collection procedures, or post-processing procedures.

A supervised training approach is necessary to generate the embeddings that are sensitive to voice quality. This is contrast to unsupervised or self-supervised approaches widely used to generate generic embeddings of speech data useful for automatic speech recognition, such as wav2vec 2.0~\cite{baevski2020wav2vec}. Embeddings trained for these tasks may not capture the disease-related speech features well or may even ignore them. To confirm this, we use the embeddings generated by wav2vec 2.0 to perform dysphonic voice classification on the SVD dataset with an SVM classifier. Its training accuracy was only 63.0\%, and the in-corpus validation accuracy was 56.76\%. We also tried transfer learning method on wav2vec 2.0 model.   That is, we use the pre-trained fine-tuned wav2vec 2.0 as a starting model and fine-tune it using labeled training data. An MLP classifier with the same structure as ours is attached to the wav2vec 2.0 model, with the input dimension modified accordingly to match the wav2vec 2.0 model. The wav2vec 2.0 model and MLP classifier are jointly trained with classification loss (learning rate is 0.001) for our task. For this setup, the training accuracy is 77.32\%, the in-corpus validation accuracy is 70.04\%, but the MEEI test accuracy is only 62.60\%, which indicates significantly worse generalizability. What's more, the dysphonic voice databases are small compared to the databases used in some unsupervised and self-supervised models. In our paper, three dysphonic voice databases are used, containing approximately 3000 speakers with only 2 hours of sustained /a/ phonation audio recordings. Whereas the wav2vec 2.0 model uses a training sample that is orders of magnitude larger.

\subsection{Comparison with Baseline Methods}

\begin{figure*}[ht]
    \centering
	\includegraphics[width=17.8cm]{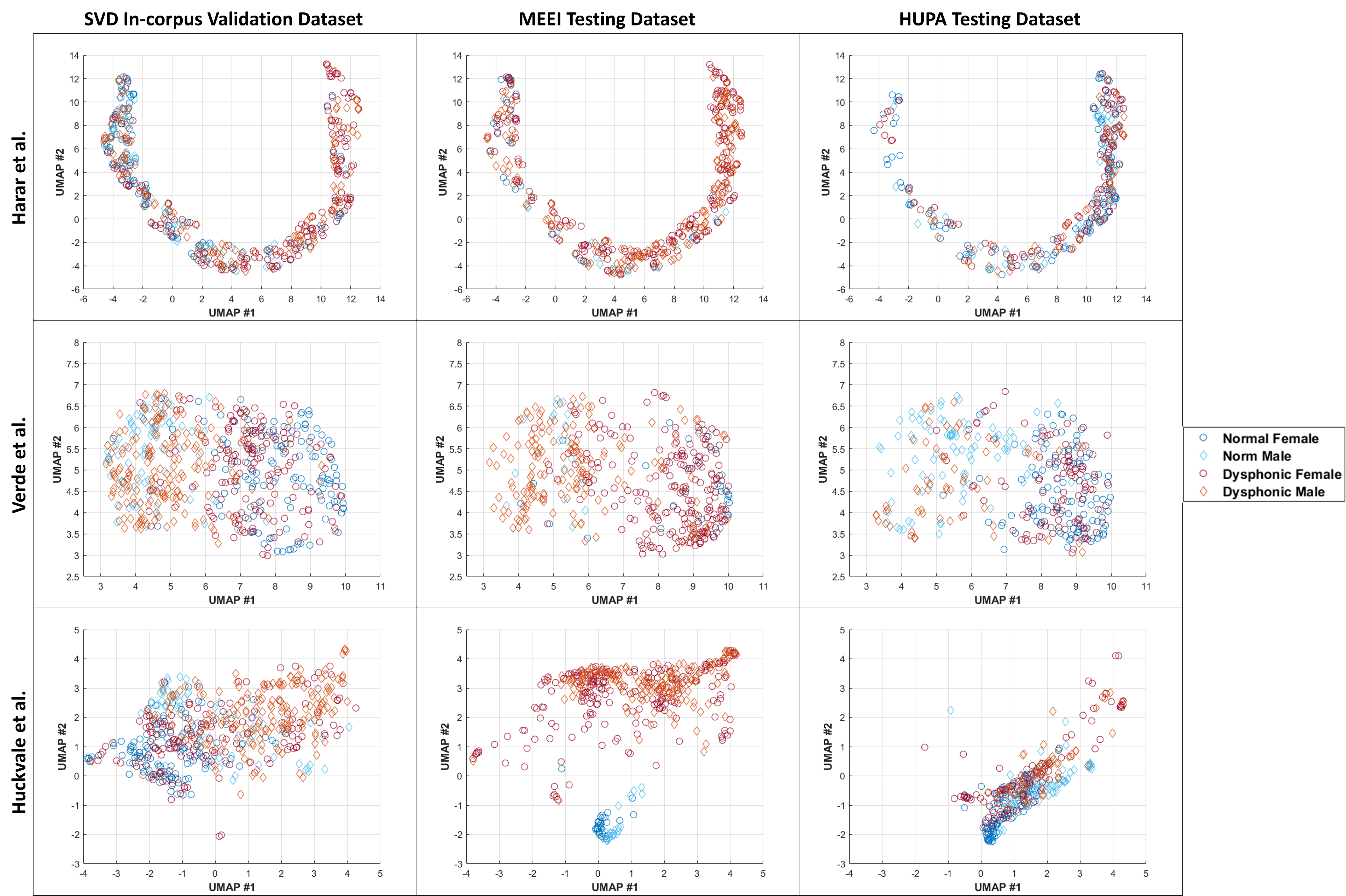}
    \caption{Embedding projection of baseline methods. Harar et al. method: the in-corpus validation and cross-corpus testing have similar projection shapes, however, the healthy samples are concentrated in a different region especially for HUPA cross-corpus testing. Verde et al. method: the dysphonic and healthy subjects are not separable. Huckvale et al. method: the embeddings UMAP projections are very different for the three corpora.}
    \label{fig:baseline}
\end{figure*}

We rebuild three baseline methods to compare with our method, and their classification accuracy and clustering metrics are shown in Tables~\ref{tab:acc} and \ref{tab:ami}. In addition, the 2-D UMAP projections of the baseline methods' embeddings are shown in Figure~\ref{fig:baseline}. For the Harar et al. method, we report SVD in-corpus validation accuracy of 68.04 ($\pm$1.11)\%, which is slightly lower than their reported in-corpus validation accuracy of 71.36\%~\cite{harar2017voice}. However, the classification accuracy is lower on cross-corpus testing, 66.14 ($\pm$2.43)\% and 49.18 ($\pm$0.80)\% for MEEI and HUPA respectively. And the clustering metrics of their method are also significantly lower than the proposed method. Furthermore, the UMAP projections of Harar et al. is not consistent between the three corpora as the healthy speech samples are concentrated in different locations.

Verde et al. reported 85.77\%\footnote{Result from~\cite{verde2018voice} Table 5, considering all parameters, method SMO.} on their SVD testing dataset by using all features with an SVM~\cite{verde2018voice}; however, their implementation only achieves SVD in-corpus validation accuracy of 62.74 ($\pm$0.91)\%, and cross-corpus testing accuracy of 70.42 ($\pm$1.83)\% and 59.76 ($\pm$1.50)\% on MEEI and HUPA respectively. Verde et al.'s method also scores low on clustering metrics, which indicates that their embeddings do not cluster by dysphonic vs. healthy. The dysphonic and healthy subjects are not separable on their UMAP projections of the embeddings.

Huckvale et al. achieve a low accuracy of 62.55 ($\pm$1.22)\% on our in-corpus validation dataset compared to their reported accuracy of 69.74\%\footnote{Result from~\cite{huckvale2021automated} Table 2, All Pathologies, Vowel, Best 1000 features.}~\cite{huckvale2021automated}. Their HUPA cross-corpus testing classification accuracy of 54.87 ($\pm$1.44)\% is lower than in-corpus accuracy. Consistent with the other two baselines, the embeddings from their method are very different across the three corpora. The OpenSMILE ComPare features show high variability across the different corpora, which coincides with Stegmann et al. findings that the OpenSMILE open-source feature package has low repeatability~\cite{stegmann2020repeatability}. 

There are many dysphonic samples that seem to be perceptually indistinguishable from healthy voices in the SVD corpus. This is also clear from Figure~\ref{fig:our} where there is some overlap between healthy and dysphonic speech samples. Only a subset of the dysphonic samples are clearly different from the healthy samples. Our results and some previous studies demonstrate that the in-corpus classification accuracy is approximately 70\% on the SVD corpus if using the complete database without filtering for specific dysphonia types~\cite{harar2017voice,huckvale2021automated}. Several previous studies discard some dysphonia sub-types that are underrepresented and cannot be fully characterized; this may lead to differing accuracy numbers in the literature compared to the works using the complete database. A consensus training and in-corpus testing splitting criteria for the SVD corpus is very helpful for future work on automated dysphonic voice detection.

Only using the classification accuracy is insufficient for evaluating different automatic dysphonic voice detection methods. A high testing accuracy does not mean the speech features set is consistent across different corpora or the deep learning model has good generalizability. The results of our work demonstrate that a method can be more comprehensively evaluated by using additional metrics that evaluate the quality of the learned embeddings (e.g., a clustering metric) and visualization for interpretability (e.g., 2-D UMAP reduction of features).

\subsection{Robustness to Degraded Recording Conditions}

Currently, many dysphonic voice/speech data collections are performed in real-world settings using mobile phone apps~\cite{nathan2019assessment,chong2021altered}; the collected recordings can be impacted by the background noise, environmental recording conditions (e.g. room acoustics), etc. These additional sources of variability can challenge the robustness of automatic dysphonic voice detection methods. To evaluate the robustness of our method and baseline methods, we simulate and apply these degradations on the original samples from our corpora. We generate several degraded in-corpus validation and cross-corpus testing datasets: one with additive background noise with a fixed SNR of 10dB (\textbf{AN}); one with simulated environmental effects via convolution with different impulse responses that mimic different microphones and environmental effects (\textbf{IR}); and a combination of the two (\textbf{AN + IR}). The implementation procedure follows the description in section IV.C. It is important to note that different background noise samples and different impulse responses were used in this test than those used during training.

The results for our method and the three baseline methods are shown in Tables~\ref{tab:acc} and \ref{tab:ami}. The in-corpus validation and cross-corpus testing accuracy of our method are consistent among the three conditions. By comparing the accuracy and clustering metric results of methods with and without data warping, it is evident that the data warping improves the generalizability of the methods across conditions. To further test this, we retrained the baseline methods with data warping and the accuracies improved across the board. Collectively, these results demonstrate that data warping is essential for good generalization across different corpora.

\section{Conclusion}

In this paper, we propose a deep learning framework for generating acoustic feature embeddings that are sensitive to vocal quality. Our framework, which contains one encoder and one MLP classifier, is jointly trained by a combined loss based on contrastive and classification loss. Data warping methods are used on the input samples to improve the robustness of our method. For the task of distinguishing the dysphonic and healthy voices, our method not only achieves good in-corpus and cross-corpus classification accuracy, but also learns consistent voice embeddings across different corpora. The proposed model also shows good generalizability on the degraded in-corpus and cross-corpus datasets. Our method outperforms the three baseline methods in all experimental conditions. A clustering metric (AMI score) and 2-D UMAP projections are used in this paper for evaluating the robustness and consistency of the voice embeddings or speech features. Our results demonstrate that this more comprehensive evaluation strategy more holistically characterizes the performance of deep learning models that learn embeddings.

In future research, the voice embeddings can be used in other disease-specific downstream tasks (e.g. Parkinson’s disease diagnosis), or in applications involving disordered speech enhancement, which can benefit patients with communication problems. Furthermore, the learned features can also be used for multi-label dysphonic voice classification. Existing corpora do not support this task due to the limited sample size and overlapping labels.

Our work also shows the benefit of data augmentation for this task. We simulated different recording devices by convolving the speech with several impulse responses. However, there is a scarcity of publicly-available resources for simulating the impulse response of recording devices (e.g. different mobile phones). A database for mobile device microphone impulse responses would be very helpful for the development of mobile health applications based on speech data in the future. Such a database can be used by developers of this technology to improve the robustness of algorithms to different recording devices and different conditions.

\section*{Acknowledgment}
This work was partially funded by the following grants: NIH-NIDCD R01DC006859 and NIH-NIDCD R21 DC019475.

\ifCLASSOPTIONcaptionsoff
  \newpage
\fi

\bibliographystyle{IEEEtran}

\bibliography{mybib}

\end{document}